\documentclass[bib,theorems,pgf]{./paper/ali-paper}

\RequirePackage{./paper/ali-macros}
\title{Public Ideological Polarization}

\author{
	Alistair Pattison%
	\footnote{Harvard University and Opportunity Insights (apattison@g.harvard.edu).}
}
\date{November 2025}

\IfFileExists{.texlabroot}{
	\pagestyle{fancy}
	\fancyhead{}
	\chead{
		\footnotesize
		\color{red!60!white}
		This pdf was compiled on \today. \\
		Please see \href{https://github.com/alipatti/spectral-radius}{\tt github.com/alipatti/spectral-radius} for the complete and most recent version of this working paper.
	}
	
}{}

\begin{document}

\maketitle

\begin{abstract}
	This paper provides a novel summary measure of ideological polarization in the American public based on the joint distribution of survey responses.
	Intuitively, polarization is maximized when views are concentrated at opposing extremes with little mass in between and when opinions are highly correlated across many issues.
	Using this measure, I show that public polarization has been increasing for the past three decades and that these changes are mostly due to increases in general disagreement, not dimensional collapse.
	Furthermore, these increases are not explained by the diverging opinions of Democrats and Republicans, nor divergence of opinions across gender, geography, education, or any other demographic divide.
\end{abstract}

\section{Introduction}
\label{sec:intro}

There is a strong sentiment among the public and the popular media that polarization is increasing \parencite{pew-2014-political-polarization, kaysen-singer-2024-movers-polarization}.
Indeed, a recent poll revealed polarization as the second-most import issue to American voters, trailing only their economic concerns \parencite{nytimes-siena-2025-registered-voter-crosstabs}.
But is the American public actually diverging in its core ideological beliefs?
Or is polarization largely superficial---an artifact of shifting attitudes rather than shifting ideology?

A large literature documents sustained growth in what has been dubbed \emph{affective} polarization---defined as rising dislike for the other side of the political aisle \parencite{iyengar-lelkes-2019-affective-polarization, boxell-gentzkow-2023-cross-country-affective, boxell-gentzkow-2017-internet-polarization, iyengar-sood-lelkes-2012-affect, lelkes-sood-iyengar-2017-hostile-audience}.
A related line of work shows that \emph{ideological} polarization---defined as divergence on policy positions and other political beliefs---has similarly increased among congresspeople and other political elites over the same time period \parencite{mccarty-poole-rosenthal-2006-polarized-america, elsas-fiselier-2023-elite-polarization-dimensions, knoll-2024-elite-polarization-boon-bane}.

However, trends in ideological polarization among the mass public are less well understood.
Are everyday Americans becoming more separated in their core ideological, moral, and political beliefs?
Some work argues that this separation has occurred only among elites with the electorate remaining similarly (un)divided \parencite{fiorina-abrams-2011-culture-war, lelkes-2016-mass-elite-polarization, glaeser-ward-2006-american-political-geography}.
Others claim meaningful growth in ideological divisions.
For example, \textcite{ojer-etal-2025-multidimensional-polarization} embed survey-respondents' opinions in a two-dimensional latent policy space and take the increasing distance between Democrats and Republicans positions as evidence of increasing party polarization.
\textcite{abramowitz-saunders-2008-polarization-myth} cite the increasing presence of consistently-liberal or consistently-conservative opinions across multiple issues.

One reason for this disagreement is the lack of a consistent definition for what constitutes ideological polarization.
This paper contributes such a definition.
I start with the observation that the mere prevalence of extreme opinions does not, by itself, constitute polarization.
What matters is opinion dispersion and---once multiple issues are considered---the structure of that dispersion across issues.
Intuitively, polarization is maximized when views are concentrated at opposing extremes with little mass in between and when opinions are highly correlated across many issues.
Formally, I summarize the joint distribution of policy opinions using the covariance matrix of survey responses and use their matrix norms as scalar indices of polarization.
My preferred index is the \emph{spectral norm} (the largest eigenvalue\footnote{
	Because covariance matrices are positive semidefinite, the spectral norm and spectral radius coincide.
	This is not the case in general.
}), which admits an intuitive interpretation as the variance of the first principal component.\footnote{
	The same framework accommodates alternative matrix norms (trace/nuclear, Frobenius), which can be interpreted as different ways of aggregating the variance of each of the principal components.
	My main results are robust to norm choice.
}
Using this measure, I find that polarization has increased on most---but not all---topics over the past three decades and that this increase starts as far back as the 1990s.

I then derive two decompositions of the spectral radius that allow me to partition both its levels and changes.
First, I decompose polarization into a term capturing the degree to which people disagree in general and a term capturing the extent to which opinions are correlated across issues.
I find that---with the notable exception of race-related issues---increases in polarization have been driven mostly by increases in general disagreement and \emph{not} by dimensional collapse.

Second, I examine the degree to which increases in polarization are explained by divergence in opinions among political parties and other demographic groups.
For example, has polarization increased simply because Democrats and Republicans disagree more strongly than they used to?
I find a nuanced answer to this question.
For example, polarization within political party is often as as high and in many cases even higher than overall polarization.%
\footnote{
	For example, Republicans are consistently more internally polarized on spending issues than the general public and Democrats are internally polarized on police and race-related issues.
}
Furthermore, I find that the driver of increasing polarization varies greatly depending on the topic.
For example, increasing polarization on race and welfare issues have been driven by between-party changes whereas increasing polarization about law-enforcement has been driven almost entirely by changes within political party.
Additionally, I find that differences of opinion between demographic groups (e.g., gender, race, geography, education, religion) in general explain a very small proportion of the observed trends.
Together, these results question existing literature that suggests increasing polarization is driven by the clustering of opinions within demographic niches \parencite{gennaioli-tabellini-2021-identity-beliefs}.

Concretely, this paper makes three contributions:
\begin{enumerate}
	\item
	      First, it motivates and develops a covariance-based measure of ideological polarization that takes into account both disagreement on individual issues and the joint distribution of opinions across many issues.
	\item
	      Second, it develops two decompositions of changes in polarization.
	      The first decomposes changes into those attributable to general disagreement and those attributable to increasing correlation of opinions across issues.
	      The second decomposes changes into within- and between-group components.
	\item
	      Third, it offers a unified portrait of mass ideological polarization over the past three decades across several topic domains by applying these methods to the University of Chicago NORC's General Social Survey \parencite{gss}.
	      I find that (1) polarization has increased and has been increasing since the 1990s, (2) these increases have been mostly driven by increases in general disagreement and \emph{not} by dimensional collapse, and (3) these increases are, in general, not explained by diverging party positions or diverging positions of demographic groups like gender, race, geography, or religion.
\end{enumerate}

The remainder of the paper proceeds as follows:
\cref{sec:theory} introduces my polarization measure and presents the statistical framework for the remainder of the paper;
\cref{sec:decompositions} derives and explains my two decompositions;
\cref{sec:gss} applies these techniques to the GSS and reports results;
\cref{sec:conclusion} concludes and outlines potential for future work.

\section{Measuring Polarization}
\label{sec:theory}

%

\subsection{On a Single Issue}

We begin by thinking about polarization on a single issue.
An initial important observation is that prevalence of extreme opinions alone is insufficient to measure what one intuitively thinks of as polarization.
For example, a question asking respondents to rate their support of slavery would get nearly universal strong opposition, yet this is not a particularly polarizing issue.
Polarization requires disagreement, i.e., spread in the distribution of responses \parencite{abramowitz-saunders-2008-polarization-myth}.

The simplest measure of a distribution's spread (and the one I proceed with for the remainder of this paper) is its variance.
On distributions with finite and bounded support---i.e., most survey questions---this closely matches one's intuition about what constitutes a highly-polarized distribution.\footnote{
	Often, polarization on a particular issue is understood as ``tail heaviness'' or the presence of a ``U'' shaped density.
	Because the distributions we're concerned about are bounded, variance captures this notion:
	As spread increases, the density has nowhere to go and must accumulate at the extremes.
}
For example, on binary a binary yes-or-no question, variance is maximized when responses are evenly split between the two options.
For a question asking one to numerically rank their agreement or disagreement with a particular statement, variance is maximized when responses are evenly split between the two extremes.
More generally, for any finite and discrete $A \subseteq \R$, the random variable with support over $A$ that has the highest variance has mass
equal to a half at the minimum and maximum values of $A$ and zero elsewhere.
One downside to using variance as the basis of our analysis is that the exact numeric values are sensitive to scale and---on their own---quite meaningless.
However, for a fixed scale, one can still get insight into the evolution of polarization over time.


\subsection{Across Multiple Issues}

In reality people hold opinions on a wide variety of issues, even within a single topic.
For example, regarding abortion, people hold separate opinions on whether abortion should be legal unilaterally, in the case of rape, in the case of incest, past the first trimester, or if the mother's health is at risk.
A good measure of polarization on abortion issues should take into account the spread of public opinion on all these scenarios.

A complete but high dimensional measure would be the distribution's covariance matrix $\Sigma$, however, this very quickly becomes uninterpretable.
A first attempt at a low-dimensional summary measure is to simply take the sum of the variances on each individual issue (i.e., the covariance matrix's trace).
This is a good start, but it fails to take into account any of the correlational structure between responses.
Indeed, much of the political science literature notes that a key feature of a ideologically-polarized society is a high-degree of correlation between opinions on separate issues, or in the words of \textcite{drutman-2020-doom-loop}, a society that is  ``locked in a zero-sum struggle along a single `us-vs-them' dimension.''

A counterintuitive, but ultimately fruitful, idea is to abstract away entirely from the specifics of our polarization setting and instead think about the goal in an abstract mathematical sense: One has some object (a covariance matrix) in a high-dimensional vector space, and we'd like some scalar measure of its ``size''.
The mathematical concept of a norm provides exactly this.
In $\R^n$, one typically uses the Euclidian ($\mathcal L_2$) norm, but unfortunately in matrix spaces, there are many choices choices for a norm with \emph{a priori} no obvious ``best'' option.
We've already met one---the trace of a matrix is indeed a norm called the \emph{trace}, \emph{Ky Fan}, or \emph{nuclear norm}.
We've already seen the trace norm's shortcomings, so I focus on two other options: the spectral norm and the Frobenius norm.

There are many equivalent ways to define the spectral norm $\norm \Sigma_2$\footnote{
	For example: as the largest singular value; as the $\mathcal L_\infty$ norm of the singular values; as the $\mathcal L_2$ operator norm.
}---the most conducive to our setting is as the largest eigenvalue of the covariance matrix.\footnote{
	This is equivalent to the spectral norm because covariance matrices are always positive semidefinite, a cone on which the spectral radius and spectral norm coincide.
}
The benefit of this definition is that the spectral norm is now easily interpretable as the variance of the responses' first principal component \parencite{anderson-1963-pca-asymptotics, jolliffe-2002-pca} and therefore combines information about the dispersion on individual issues with how much the dispersion projects onto a single dimension.
Indeed, for a fixed trace, the spectral norm is maximized when the covariance matrix has rank 1, i.e., all variance is explained by a single factor.
I explore this interpretation further in \cref{sec:trace-concentration}.

The Frobenius norm $\norm{\Sigma}_F$ also can be understood in terms of principal components:
If $\lambda_1, \ldots, \lambda_p$ are the eigenvalues of $\Sigma$ (equivalently, the variances of each of the principal components), the Frobenius norm can be written as $\sqrt{\lambda_1^2 + \cdots + \lambda_p^2}$, i.e., the Euclidean norm of the eigenvalue vector.
In some sense, this is a ``smoother'' summary of $\Sigma$'s eigenvalues than the spectral norm which simply picks out the largest.

\begin{figure}
	\caption{Comparison of Matrix Norms}
	\label{fig:norm-comparison}
	\begin{minipage}{.5\textwidth}
		\includegraphics[width=\textwidth]
		{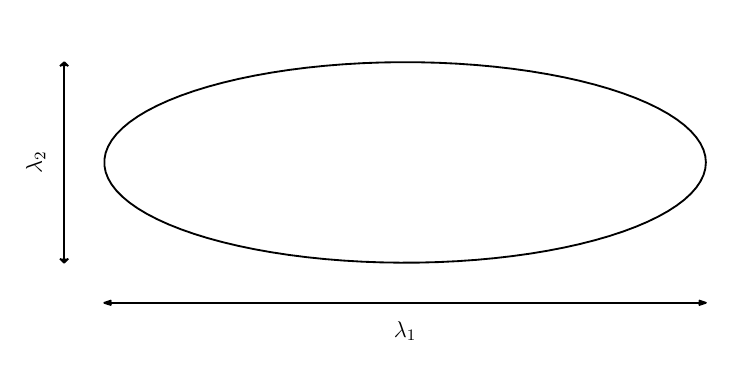}%
	\end{minipage}%
	\hspace{.25in}
	\begin{tabular}{r l | c | c}
		\multicolumn{2}{c|}{Norm} & Expression       & $p$                                           \\
		\hline
		Spectral                  & $\norm \Sigma_2$ & $\max(\lambda_1, \lambda_2)$       & $\infty$ \\
		Frobenius                 & $\norm \Sigma_F$ & $\sqrt{\lambda_1^2 + \lambda_2^2}$ & $2$      \\
		Nuclear                   & $\norm \Sigma_*$ & $\lambda_1 + \lambda_2$            & $1$      \\
	\end{tabular}
	\notes{
		The left-hand panel shows the ellipse $\set{ \mathbf x \trans \Sigma \mathbf x : \norm{\mathbf x} = 1 }$ induced by the matrix $\Sigma \in \R^{2 \times 2}$ with eigenvalues $\lambda_1$ and $\lambda_2$.
		The right-hand panel shows how our three matrix norms can be expressed as $p$-norms of the the matrix's eigenvalue vector, $(\lambda_1, \, \lambda_2) \trans$.
		The spectral norm measures the maximum radius of the ellipse, the nuclear norm measures the sum of all radii, and the Frobenius norm provides a smooth in-between.
	}
\end{figure}

\begin{figure}
	\centering
	\caption{Responses to Two Related Survey Questions}
	\label{fig:gss-example}
	\includegraphics{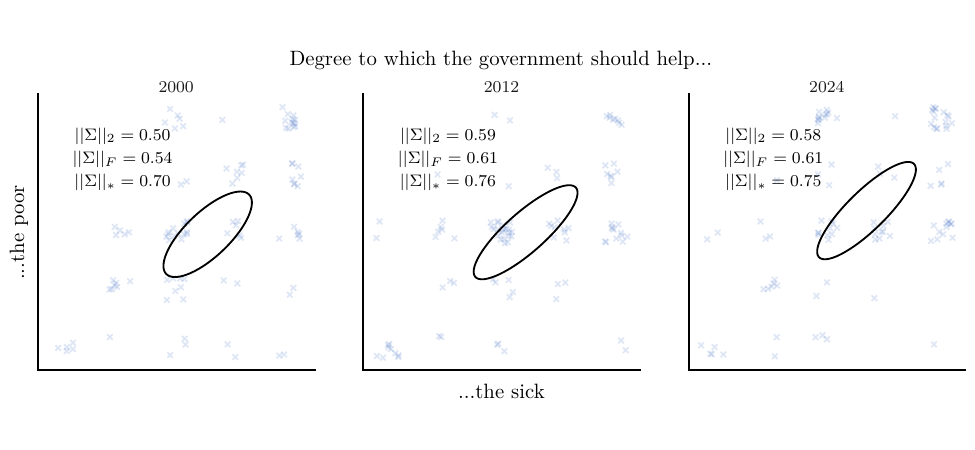}
	\notes{
		This figure plots 100 yearly responses from the General Social Survey on two questions regarding the responsibility of government to help those who are sick ($x$-axis) or who are poor ($y$-axis).
		Superimposed is the yearly covariance ellipse $\set{ \mathbf x \trans \Sigma \mathbf x : \norm{\mathbf x} = 1 }$.
		The position of each point has been slightly jittered for plotting.
		From 2000 to 2012, the distribution of responses stretches causing an increase in all three of our measures of polarization.
		From 2012 to 2024, the entire distribution shifts upwards and to the right, causing an increase in the number of extreme responses.
		But the spread remains constant and our polarization measures do not budge.
	}
\end{figure}

\cref{fig:norm-comparison} summarizes how each norm relates to the eigenvalues of the covariance matrix, and \cref{fig:gss-example} provides an illustration of each of these norms applied to survey responses on two issues from the General Social Survey \parencite{gss}.
I elect to use the spectral norm for the remainder of the paper due to its simple interpretation as the variance of the first principal component, but the overall trends qualitatively hold for other norm choices.

\subsection{Statistical Formalization and Notation}
\label{sec:formalization}

I now formalize our data setup and estimand, present an estimator, and prove some basic properties about it.
In particular, I assume that I observe i.i.d. samples $\mathbf x_1, \ldots, \mathbf x_n \sim \mathcal F$ with $\mathbf x_i = \paren{x_{i1}, \ldots, x_{ip}}$, drawn from some underlying distribution of political opinions $\mathcal F$.
Each $x_{ij}$ is the opinion of individual $i$ on issue $j$.
Notably, I make no assumptions about the distribution $\mathcal F$ from which the responses are drawn except that it has a finite covariance matrix $\Sigma = \Var(\mathbf x_i)$.
Let $\lambda_1, \ldots, \lambda_p$ be the eigenvalues of $\Sigma$ ordered from largest to smallest.
Then, the estimand of interest is $\rho = \norm{\Sigma}_2 = \lambda_1$, the spectral radius of $\Sigma$.

My estimator is quite straightforward:
First, I define the sample covariance matrix $\widehat S_n \ceq \frac 1n \sum_i \mathbf x_i \mathbf x_i \trans$ and corresponding sample eigenvalues $\hat \lambda_1, \ldots, \hat \lambda_p$.
Then, my estimator is simply $\hat \rho = \| \widehat S_n \|_2 = \hat \lambda_1$.
This turns out to be okay (proof follows from a straightforward application of the delta method; see \cref{sec:deferred-proofs} for details):

\begin{proposition}
	The spectral radius of the sample covariance matrix is a consistent and asymptotically-normal estimator of the population spectral radius.
\end{proposition}

\subsection{Relationship to Spread in a Latent Ideology Model}

My setup is intentionally model-free, however, it does play nicely with a one-dimensional latent model (for example, a latent left/right political ideology).
In particular, the spectral radius picks up increases in spread within the latent distribution.
The following proposition formalizes that notion:

\begin{proposition}
	\label{thm:one-dim-implies-multiple-dim}
	Let $y \sim \mathcal G$ be a person's latent one-dimensional ideological position modeled as random variable with positive variance $0 < a = \Var(y) < \infty$.
	Let $\beta  = \paren{\beta_1, \ldots, \beta_p} \in \R^p$ be the sensitivity of each of issue to this latent ideology such that one's revealed policy positions are
	\begin{equation}
		x_{ij} = \beta_j y + e_i;
		\hspace{1in}
		\Var(e_1, \ldots, e_p) = \Gamma \in \R^{p \times p}.
	\end{equation}
	Define $\Sigma = \Var(\mathbf x_i)$ and $r = \norm{\Sigma}_2$.
	Then the following hold:
	\begin{enumerate}
		\item $r$ is non-decreasing in $a$;
		\item if $\beta$ nontrivially projects onto the principal eigenspace of $\Gamma$, then $r$ is strictly increasing in $a$;
		\item if $a > \norm{\Gamma}_2$, then $r$ is strictly increasing in $a$.
	\end{enumerate}
\end{proposition}

Note that this proposition is quite flexible: I make no assumptions about the latent distribution other than it has a finite second moment.
Additionally, the error term is also allowed to take any form---in particular it can have complex non-diagonal covariance structure.
See \cref{sec:deferred-proofs} for proof.

\section{Decompositions}
\label{sec:decompositions}

I now develop the two decompositions introduced in \cref{sec:intro}.

\subsection{General Disagreement and Cross-Issue Correlation}
\label{sec:trace-concentration}

The fact that the spectral norm of the covariance matrix can be understood as the variance of the first principal component means that $\norm \Sigma _2$ admits a straightforward multiplicative decomposition as the total variance ($\tr \Sigma$) times the proportion of variance explained by the first principal component ($\lambda_1 / \tr \Sigma$):
\begin{equation}
	\norm{\Sigma}_2
	\ = \ \lambda_1
	\ = \ \tr \Sigma \cdot \lambda_1 / \tr \Sigma.
\end{equation}
I call the first component the total variance and the second component the ``spectral concentration''.

By itself, this decomposition is somewhat useful in telling us how much variation projects onto a single dimension, but it's real utility comes in unpacking the source of the difference between two distributions' polarization.
For example, suppose that there are two distributions $\mathcal F_0$ and $\mathcal F_t$ with covariance matrices $\Sigma_0$ and $\Sigma_t$ respectively.
Then, I can decompose the percentage difference in their polarizations as the change in total variance times the change in spectral concentration:
\begin{equation}
	\label{eq:percent-change-trace-concentration}
	\frac{\norm{\Sigma_1}_2}{\norm{\Sigma_0}_2}
	\ = \ \frac{\tr \Sigma_1}{\tr \Sigma_0} \cdot \frac{\lambda_{0,1} / \tr \Sigma_0}{\lambda_{t,1} / \tr \Sigma_t}
\end{equation}

\begin{figure}
	\centering
	\caption{Same Increase in Spectral Radius, Two Different Reasons}
	\label{fig:stretch-v-scale}
	\includegraphics{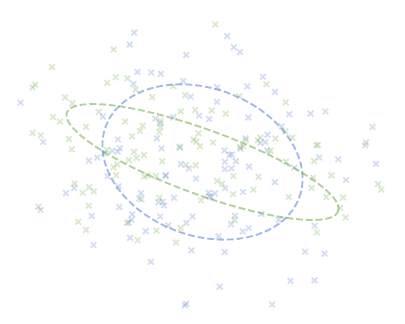}%
	\hspace{.5in}
	\includegraphics{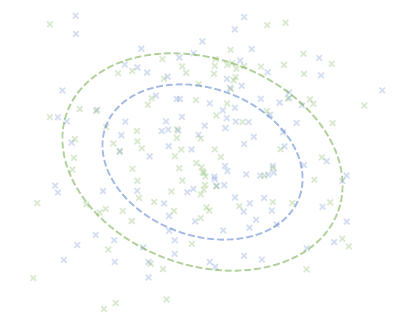}%
	\notes{
		The spectral norm can be multiplicatively decomposed into the product of the spectral concentration (i.e., the proportion of variance explained by the first principal component) and the trace of the covariance matrix (i.e., the sum of the individual variances).
		In this figure, the blue distributions are the same in both panels and the green distributions share the same larger spectral radius.
		However, the reasons for this increase are quite different.
		In the left-hand panel, the larger norm is entirely due to an increase of the relevance of the first principal component with the total variance of the two distributions held constant.
		In the right-hand panel, the difference is entirely due to the total variance with a fixed spectral concentration.
	}
\end{figure}

\Cref{fig:stretch-v-scale} illustrates this idea, and in \cref{sec:trace-concentration-decomp} discusses results from applying this technique to survey data to decompose changes in the polarization of the American public over time.

\begin{figure}
	\centering
	\caption{Polarization Within Subgroups Can Be Lower or Higher Than Overall Polarization}
	\label{fig:between-group-sign}
	\includegraphics{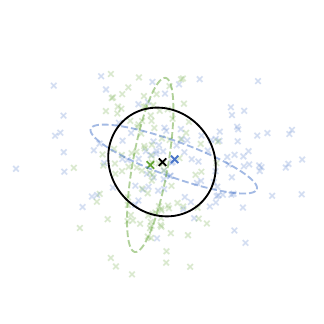}%
	\includegraphics{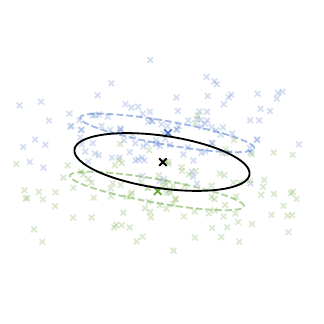}%
	\includegraphics{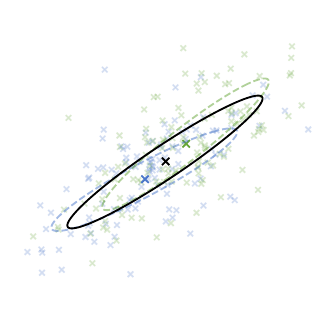}
	\notes{
		The sign and magnitude of of the between-group polarization $\rho_b$ can vary greatly depending on the means and covariance structure of the groups.
		In the left-hand panel, the individual subgroups (dashed) are distributed such that the within-group polarization is higher than the pooled (black) polarization, i.e., $\rho_b < 0$.
		In the center panel, the within-group principal eigenvectors are orthogonal to the between group disagreement, so $\rho_b = 0$.
		The rightmost panel shows a scenario where the within- and between-group polarizations align to yield $\rho_b > 0$.
	}
\end{figure}

\subsection{Between vs. Within Groups}
\label{sec:between-v-within}

The following section explores how polarization can come from both disagreement \emph{within} a particular group (e.g. disagreement among Democrats about immigration) and disagreement \emph{between} groups (e.g. Democrats and Republicans hold, on average, quite different opinions about law enforcement).
To start, I amend our statistical setup to allow for such groupings.
In particular, let
\begin{equation}
	z_i \sim \operatorname{Multinomial}(p_1, \ldots, p_G);
	\hspace{1in}
	p_1 + \cdots + p_G = 1
\end{equation}
be an integer indicating the group to which individual $i$ belongs.
(In particular, the group $z_i$ may be correlated with opinions $x_{ij}$.)
Then, the multidimensional analogue to he law of total variance \parencite[e.g.]{anderson-2003-multivariate} allows us to decompose the covariance matrix of $\mathbf x_i = (x_{i1}, \ldots, x_{ip})$ into a within- and between-group component:
\begin{equation}
	\begin{aligned}
		\label{eq:within-between-decomposition-of-sigma}
		\Sigma
		 & = \Var(\mathbf x_i)                                                         \\
		 & = \E{\Var(\mathbf x_i \given z_i)} + \Var\paren{\E{\mathbf x_i \given z_i}} \\
		 & = \underbrace{\sum_g p_g \Sigma_g}_{\substack{\text{within-group}           \\ \text{components}}} \ + \ \underbrace{\sum_g p_g \paren{\mu_g - \mu}\paren{\mu_g - \mu} \trans}_{\text{between-group component}}
	\end{aligned}
\end{equation}
where $\mu_g = \E{\mathbf x_{i} \given z_i = g}$ are the average opinions of group $g$, $\Sigma_g = \Var(\mathbf x_i \given z_i = g)$ is the group's covariance matrix, and $\mu = \E{\mathbf x_i}$ is the unconditional mean.
Unfortunately, the spectral norm is not linear, so this does not immediately yield a within- and between-group decomposition of $\norm \Sigma _2$.

Letting $\Sigma_w$ and $\Sigma_b$ denote the within- and between-group components of the decomposition in \cref{eq:within-between-decomposition-of-sigma}, the triangle inequality lets us write $\norm{\Sigma}_2$ as the sum of $\norm{\Sigma_w}_2$ and $\norm{\Sigma_b}_2$, less an additional slack term $s_b \geq 0$ that captures the tightness of the triangle inequality for this particular pair of matrices:
\begin{equation}
	\label{eq:norm-sigma-decomposition-1}
	\norm{\Sigma}_2 = \norm{\Sigma_w}_2 + \norm{\Sigma_b}_2 - s_b.
\end{equation}
In general, the interpretation of $s_b$ varies depending on vector space and the norm being used,\footnote{
	For example, with the $\mathcal L_2$ norm on $\R^n$, the slack is related to the angle between the two vectors.
} but in our spectral-norm setup, $s_b$ captures the degree to which the principal eigenvectors of $\Sigma_w$ and $\Sigma_b$ align, with $s_b$ vanishing precisely when the principal eigenvectors are collinear.
In our setting, this is means that $s_b$ captures the difference in the direction of polarization between $\Sigma_w$ and $\Sigma_b$:
Small $s_b$ means that groups disagree along a similar axis amongst themselves as the disagreement between groups.
Large $s_b$ means that the directions of polarization are different.

However, one is often concerned not just with difference in the direction of polarization between $\Sigma_b$ and $\Sigma_w$, but the differences in the directions of polarization among the groups themselves.
Are Democrats internally polarized along the same axes as Republicans?
To get at this, I can further decompose $\norm{\Sigma_2}_2$ into the norms of its individual group-specific covariance matrices plus another slack term $s_b \geq 0$:
\begin{equation}
	\norm{\Sigma}_2 = \sum_i p_i \norm{\Sigma_i}_2 + s_w - \norm{\Sigma_b}_2 - s_b.
\end{equation}
Here, $s_w$ captures the degree to which each group is polarized on a different set of issues in the same way that $s_b$ in \cref{eq:norm-sigma-decomposition-1} above captured how the direction of polarization differed between $\Sigma_w$ and $\Sigma_b$.
For simplicity of presentation, I typically combine both slack terms and the between-group polarization into a single ``between-group'' category:
\begin{equation}
	\label{eq:norm-sigma-decomposition-final}
	\norm{\Sigma}_2
	\ = \ \underbrace{\sum_i p_i \norm{\Sigma_i}_2}_{\rho_{\text{within}}} + \underbrace{\norm{\Sigma_b}_2 - s_w - s_b}_{\rho_{\text{between}}}.
\end{equation}

The payoff from all this math is that both terms in \cref{eq:norm-sigma-decomposition-final} have an intuitive interpretation:
\begin{itemize}
	\item
	      $\rho_{\text{within}}$ measures the extent to which each group is polarized amongst themselves and is simply the weighted average of the group-specific polarizations.
	\item
	      $\rho_{\text{between}}$ measures the extent to which there is polarization across groups. The norm of the between-group covariance matrix captures increased polarization coming from different groups holding different positions.
	      The slack terms accounts for the fact that overall polarization is diminished if different groups are divided on different issues.
	      Note that $\rho_b$ is \emph{not necessarily positive}—it can be (and often is the case) that the differing directions of polarization in individual groups ``cancel out'' the between group polarization.
	      \Cref{fig:between-group-sign} visualizes how between-group differences can amplify, offset, or leave unchanged overall polarization depending on the axis along which each subgroup is polarized.
\end{itemize}

Similarly to the previous section, the utility of this decomposition comes largely from its ability to partition \emph{changes} into those due to within-group vs between-group increases in polarization.
In particular, \cref{sec:party-decomp} presents results where I fix either the within-or between group component to its initial value and let only the other component play out over time.

\section{Application in the General Social Survey}
\label{sec:gss}

I now apply these techniques to the General Social Survey---a bi-annual poll of the American public on a broad variety of social, political, and ideological topics administered by NORC at the University of Chicago \parencite{gss}.
I restrict to questions that have been asked repeatedly over the past three decades and manually categorize questions into eight topics: abortion, affirmative action, free speech, government spending, police and justice, race, sex and birth control, and welfare.
For each category, \cref{sec:gss-questions} enumerates the selected survey questions, their text, and the possible responses.

To encode survey responses as numeric values to which I can apply my polarization measure, I map the minimum response to -1, the maximum response to +1, and the remaining responses evenly-spaced between those two values.
For example, binary questions are encoded as $\pm 1$ and questions with three options are mapped into $(-1, 0, +1)$.
I only include questions that are binary or admit some ordinal structure (e.g. rating one's agreement with a statement).%
\footnote{A future version of this working paper will explore the impact of different methods of numerically encoding survey responses (e.g., using quantiles).}
To deal with missingness (arising from certain people receiving only a subset of survey questions and people electing not to respond), I estimate each entry of the covariance matrix using all pairs for which both questions contain a response.%
\footnote{
	The is equivalent to a missing-at-random assumption, which I will evaluate in a future version of this working paper.
	However, to affect any of our trends, this bias would have to dynamically evolve over time.
	This also does create the possibility of a non-PSD covariance estimate, but the probability of this vanishes as $n \to \infty$.
}
I use the provided \texttt{WTSSPS} survey weight.

\subsection{Overall Trends}
\label{sec:overall-trends}

\begin{figure}
	\centering
	\caption{Polarization Over Time by Issue Category}
	\label{fig:gss-time-series}
	\input{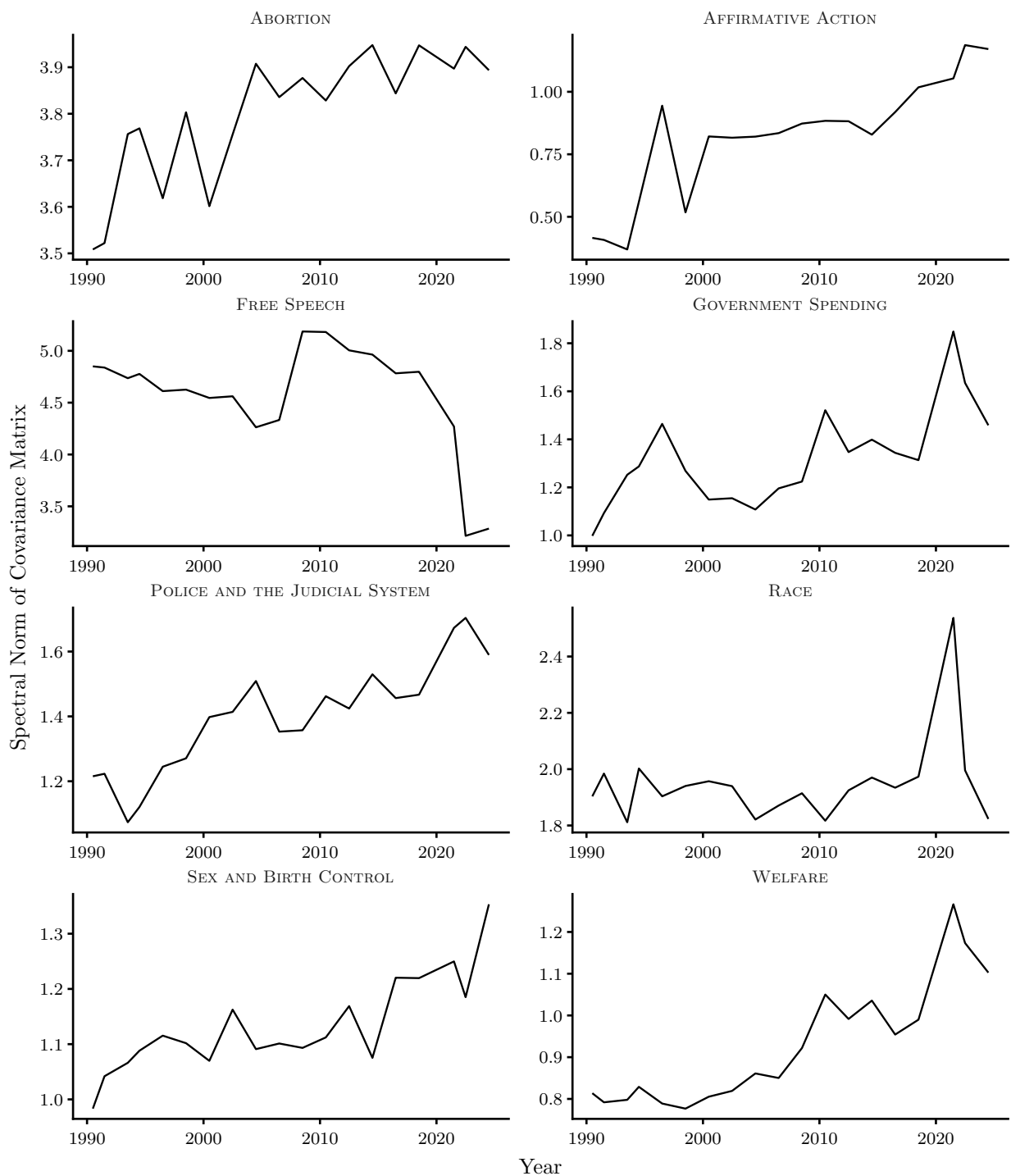}
	\notes{
	The $x$-axis shows year; the $y$-axis shows the spectral radius of the covariance matrix for responses to GSS survey questions in that year.
	\Cref{sec:gss-questions} enumerates the questions included in each category.
	Bootstrap standard errors leveraging asymptotic normality are forthcoming in a future version of this working paper.
}
\end{figure}

\cref{fig:gss-time-series} shows how polarization---as measured by the spectral radius---evolves over time for the eight previously-mentioned topics.
I find steady increases in polarization on all topics except free speech and race.
Free speech issues exhibit steady decline in polarization throughout the 1990s and 2000s before spiking in the wake of the 2008 financial crisis.
Polarization then remains high throughout the 2010s before plummeting again during the COVID-19 pandemic.
On race issues, polarization remains flat throughout our entire period with the exception of an enormous spike in polarization during the early 2020s.
In addition to the general upward trend, spending polarization exhibits spikes during the mid 1990s, the 2008 financial crisis, and the COVID-19 pandemic.

\subsection{Dimensional Collapse or Increase in General Disagreement?}
\label{sec:trace-concentration-decomp}

In \cref{sec:trace-concentration}, I demonstrated a decomposition of the spectral radius into the trace of the covariance matrix (summarizing total disagreement and ignoring correctional structure) times the share of variance explained by the first principal component (summarizing the degree to which opinions are correlated across issues).
I now explore which of these two components is driving the observed trends.
Are increases in polarization stemming from people becoming increasingly one dimensional in their opinions, or are people simply disagreeing more in general?

\begin{figure}
	\centering
	\caption{Changes due to Spectral Concentration Versus Total Variance}
	\label{fig:gss-spectral-concentration}
	\input{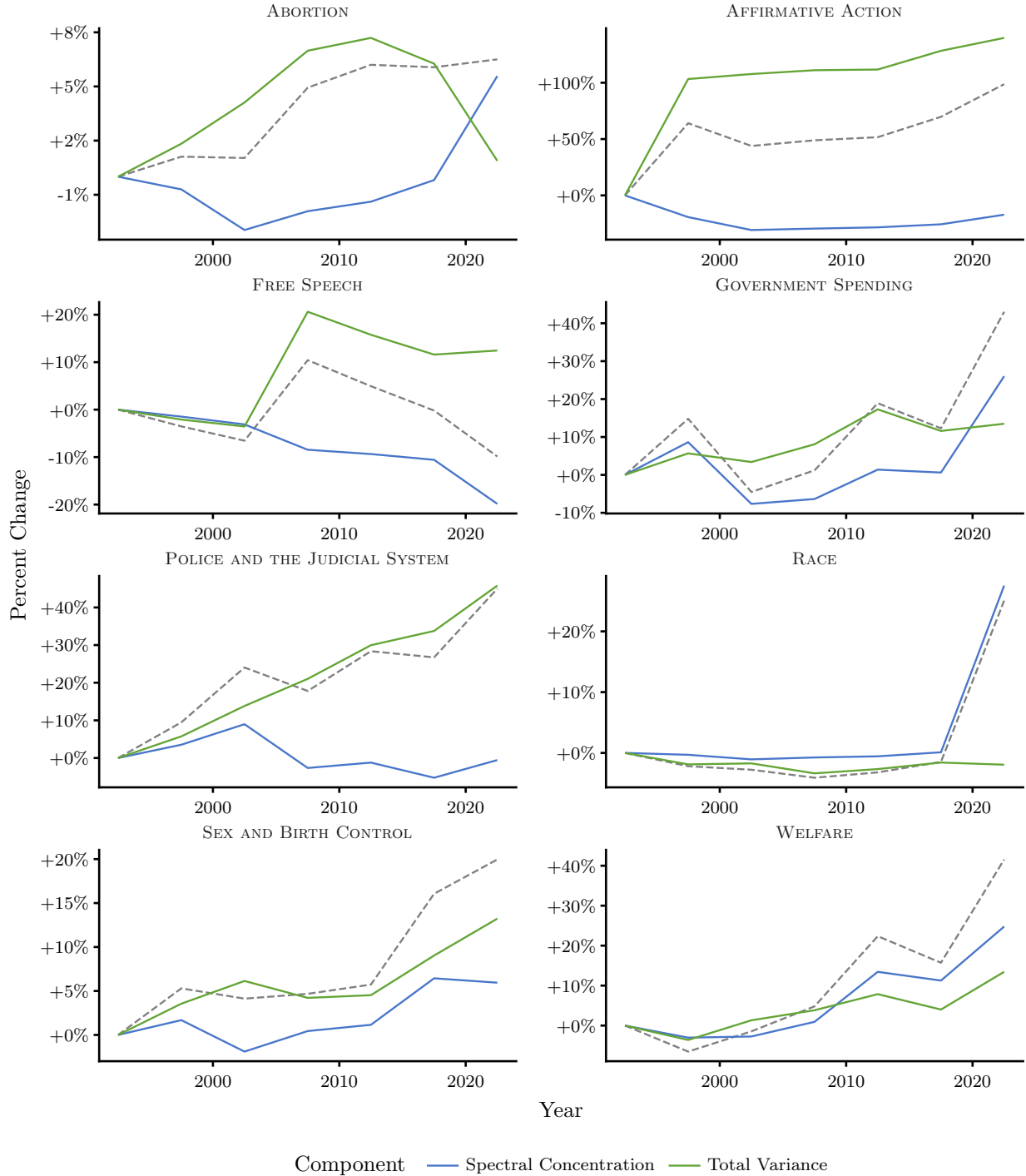}
	\notes{
		The gray line shows the true observed change in polarization.
		The blue line shows the ``dimensional-collapse-only'' counterfactual holding total variance to its original value.
		The green line shows the ``general-disagreement-only'' counterfactual by fixing spectral concentration at its original value.
		The two counterfactuals multiply to produce the observed change.
		I pool to five-year bins for additional power.
	}
\end{figure}

To determine the proportion of change attributable to dimensional collapse, I fix the total variance component of \cref{eq:percent-change-trace-concentration} to its initial value and let only the spectral concentration evolve over time. To see the change attributable to increase in total variance, I fix the spectral concentration to its initial value and let only the total variance evolve. These two counterfactuals multiply together to equal the true observed change. \cref{fig:gss-spectral-concentration} illustrates the results.

In general, I find that most trends are driven by increases in total variance with spectral concentration sometimes even decreasing over time (e.g. speech, police).
The most obvious exceptions to this trend are (1) race issues, where increases in spectral concentration explain the entire spike in the early 2020s, and (2) abortion issues, which exhibit a ``U'' shape with total variance increasing from 1990 until around 2010 before decreasing sharply from 2010 until the present.

\subsection{Within or Between Political Parties?}
\label{sec:party-decomp}

\Cref{sec:between-v-within} developed a way to decompose polarization into portions attributable to within- and between-group disagreement.
I now apply this technique to the GSS.

As a first step, \cref{fig:gss-within-party} shows the levels and trends of polarization \emph{within} each political party.
I counterintuitively find that within-party polarization is typically comparable to---and in some cases higher than---polarization in the general population.
Democrats are typically more polarized on police, race and affirmative action and are united on welfare and spending issues.
Republicans are polarized on spending and sex and united on policing, race, and affirmative action.%
%
\footnote{
	Party is endogenous to one's political and ideological beliefs, so results from this section include dynamic sorting of people into party.
	I also present results by self-reported ideology (liberal v. conservative) in \cref{sec:other-decompositions}.
}

I then decompose changes in polarization into those attributable to within- and between-party trends;
\cref{fig:gss-within-v-between} shows these results.
I find that the breakdown varies across topic:
Trends in affirmative action, free speech, police polarization are driven entirely by within-party changes.
Trends in polarization on race and welfare issues are driven entirely by between-party changes.
Interestingly, spending polarization exhibits a ``switch'': Volatility in the 1990s and in the wake of the 2008 financial crisis stemmed from within-party disagreement, but the COVID-19 spending polarization bump was driven by between-party differences.

\subsection{What About Gender? Age? Race? Religion?}
\label{sec:gss-other-groups}

\begin{figure}
	\centering
	\caption{Polarization Within Political Party}
	\label{fig:gss-within-party}
	\input{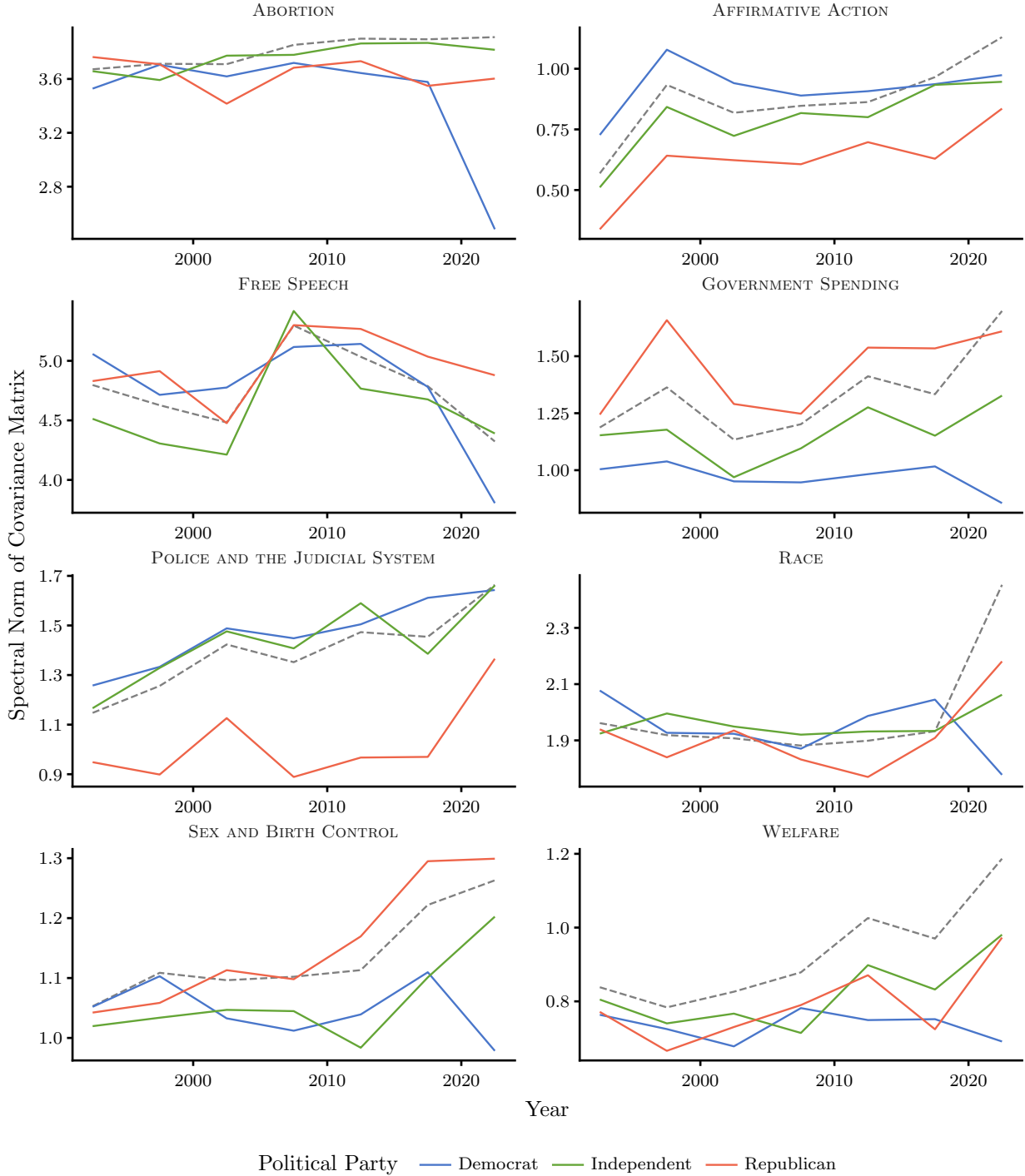}
	\notes{
		They gray line shows overall polarization.
		The colored lines show polarization within self-identified political party.
		I pool to five-year bins for additional power.
	}
\end{figure}

\begin{figure}
	\centering
	\caption{Changes From Within- Versus Between-Group Polarization}
	\label{fig:gss-within-v-between}
	\input{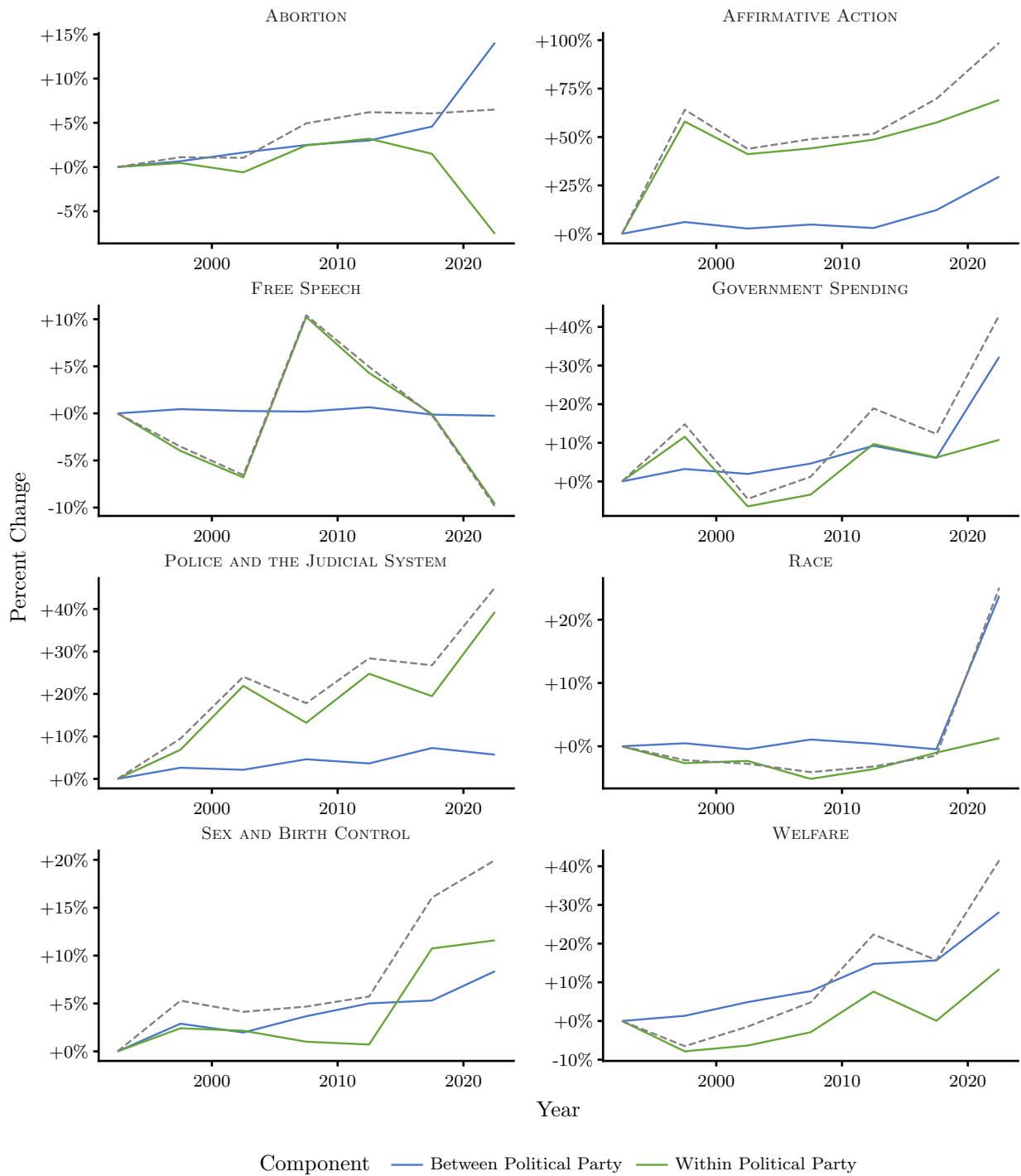}
	\notes{
		The gray line shows the true observed change in polarization.
		The blue line shows the ``between-party-only'' counterfactual holding the within-group polarization fixed.
		The green line shows the ``within-group-only'' counterfactual holding the between-group polarization fixed.
		The two counterfactuals add to produce the observed change.
		I pool to five-year bins for additional power.
	}
\end{figure}

If the divergence of political parties is not entirely to blame for increasing polarization, are there other demographic characteristics that explain the increases?
For example, is increasing polarization on race issues being driven by divergence of the opinions of white and black people?
Are increases in abortion polarization being driven by a divergence of men's and women's views?
It appears that in the vast majority of cases, the answer is no---most trends stem from within-group changes, not divergence of opinions across groups.
Two notable exceptions are that education and religion explain a large chunk of the increase in abortion polarization.
See \cref{sec:other-decompositions} for analogues to \cref{fig:gss-within-party,fig:gss-within-v-between}, replacing political party with race, sex, religion, education, geography, age, and other demographics.

\section{Conclusion}
\label{sec:conclusion}

In this paper, I developed index of mass ideological polarization based on the joint distribution of people's opinions.
Concretely, I computed the covariance matrix of responses to various survey questions about views on policy and ideology and took my measure to be the matrix's spectral radius.
This captures both the degree to which people disagree on individual issues \emph{and} the between-issue correlation of opinions (\cref{sec:theory}).

Next, I presented two decompositions that explored not just whether polarization has increased, but \emph{why}.
The first decomposition partitioned changes into a term capturing the degree to which opinions have become increasingly one-dimensional and a term capturing the strength of disagreement in general.
The second decomposition partitioned changes into within- and between-group components (\cref{sec:decompositions}).

Finally, I applied these methods to three decades of survey data, producing an evolving portrait of ideological polarization in the American public (\cref{sec:gss}).
I found that polarization has increased on most topics since the early 1990s (\cref{sec:overall-trends}), driven mostly by increases in general disagreement rather than increased cross-issue opinion correlation (\cref{sec:trace-concentration-decomp}) and by increases in disagreement within groups rather than between them (\cref{sec:party-decomp,sec:gss-other-groups}).
Taken together, this evidence challenges accounts that increasing polarization is due to the ideological shift of entire demographic groups.

\subsection{Limitation vs and Future Work}

This paper leaves significant room for future work.
First, there are many topics not covered by the GSS that I would like to explore.
For example, it asks no long-running questions about immigration.
Repeating this analysis with other data like the \textcite{anes} could provide additional insight.

Second, the repeated cross-section structure of my data means that the within- versus between-party analysis is effectively conditioning on an endogenous characteristic and may reflect dynamic sorting into groups rather than true changes \parencite{abramowitz-webster-2016-negative-partisanship}.
Repeating the analysis with a panel data set would allow me to fix political party (or any other covariate) at some initial $t=0$, and then allow opinions to evolve within those static groups.%
\footnote{The GSS and Pew Research do provide panel survey data sets, although their scope is fairly small and both panels cover a far shorter of a time period than the corresponding repeated cross-section.}

Third, future work could explore the relationship between polarization of political elites (e.g. congresspeople) and the polarization of their constituents.
Do polarized electorates produce polarize politicians?
Similarly, do polarized leaders produce polarized constituents?
One could also extend the decompositions of \cref{sec:decompositions} to attribute trends in congressional ideological polarization to the turnover of elected officials versus the changing opinions of incumbents.
Are lifetime politicians becoming more polarized in their views, or is congressional polarization a symptom of new representatives with strong opinions?

Fourth, how does ideological polarization relate to public discourse?
Does the ideological polarization detected by the spectral radius correlate over time with the prevalence of an issue in the media?
Does highly-polarized discussion on Twitter (in the affective sense) correlate with the ideological polarization of the broader public?

Finally, how does ideological polarization vary at a fine geographic level?%
\footnote{
	This analysis would be possible using the \href{https://gss.norc.org/content/dam/gss/get-documentation/pdf/other/ObtainingGSSSensitiveDataFiles.pdf}{restricted access GSS data with granular geographic identifiers}.
}
What are the hot spots for disagreement?
Where is there an ideological monoculture?
What causes this variation?
Producing such fine estimates would require reckoning with finite-sample upward bias in the naïve spectral radius estimator for extremely small cells, among other statistical issues.%
\footnote{
	Putting a prior on the covariance matrix (or its diagonalizing operator) would reduce this bias at the cost of imposing an assumption that people disagree similarly (or on similar issues) across the whole country.
	A Markov field would relax that assumption and allow the direction of disagreement to vary smoothly over space.
}
Assuming that such problems can be tackled, do we observe the same increase in polarization in the places most exposed to Chinese import competition as found by \textcite{autor-dorn-hanson-majlesi-2020-importing-political-polarization}.
What about internet activity?
Previous work on its impact have produced conflicting results \parencite{boxell-gentzkow-2017-internet-polarization, lelkes-sood-iyengar-2017-hostile-audience}.
What about other shocks?
Does exposure to immigration raise or lower race-related polarization among the native-born population?

Altogether, I hope that this work provides a meaningful contribution to the discourse around political polarization and serves as a useful foundation for future analysis.

\newpage

\printbibliography

@misc{gss,
        title = {General Social Survey},
        author = {Davern, Michael and Bautista, Rene and Freese, Jeremy and Herd
                  , Pamela and Morgan, Stephen L.},
        year = {2025},
        month = {1},
        publisher = {NORC at the University of Chicago},
        url = {https://gssdataexplorer.norc.org/gsscite},
}

@misc{anes,
        title = {American National Election Studies},
        author = {{American National Election Studies Program}},
        year = {2021},
}

@misc{nytimes-siena-2025-registered-voter-crosstabs,
        title = {Registered Voter Cross-Tabs},
        author = {The New York Times and Siena Research Institute},
        shortauthor = {Times and Siena},
        year = {2025},
        month = {10},
        howpublished = {Interactive data release: “Times/Siena Poll – Registered
                        Voter Crosstabs”},
        url = {
               https://www.nytimes.com/interactive/2025/10/02/polls/times-siena-poll-registered-voter-crosstabs.html
               },
}

@report{pew-2014-political-polarization,
        author = {{Pew Research Center}},
        title = {Political Polarization in the American Public},
        year = {2014},
        institution = {Pew Research Center},
        url = {
               https://www.pewresearch.org/politics/2014/06/12/political-polarization-in-the-american-public/
               },
        urldate = {2025-11-23},
        note = {June 12, 2014},
}

@article{kaysen-singer-2024-movers-polarization,
        author = {Ronda Kaysen and Ethan Singer},
        title = {Millions of Movers Reveal American Polarization in Action},
        journal = {The New York Times},
        date = {2024-10-30},
        url = {
               https://www.nytimes.com/interactive/2024/10/30/upshot/voters-moving-polarization.html
               },
}

@article{golub-1973-eigenvalues,
        author = {Golub, Gene H.},
        title = {Some Modified Matrix Eigenvalue Problems},
        journal = {SIAM Review},
        volume = {15},
        number = {2},
        pages = {318-334},
        year = {1973},
        doi = {10.1137/1015032},
        URL = { https://doi.org/10.1137/1015032 },
        eprint = { https://doi.org/10.1137/1015032 },
}

@misc{tao-2010-eigenvalue-blog,
        author = {Terence Tao},
        title = {254A, Notes 3a: Eigenvalues and sums of Hermitian matrices},
        note = {Blog post, January 12, 2010},
        year = {2010},
        month = {1},
        url = {
               https://terrytao.wordpress.com/2010/01/12/254a-notes-3a-eigenvalues-and-sums-of-hermitian-matrices/
               },
}

@article{weyl-1912-inequalities,
        author = {Hermann Weyl},
        title = {The asymptotic distribution law of the eigenvalues of linear
                 partial differential equations (with an application to the
                 theory of cavity radiation)},
        journal = {Mathematische Annalen},
        volume = {71},
        number = {},
        pages = {441--479},
        year = {1912},
        month = dec,
        doi = {10.1007/BF01456804},
        url = {https://doi.org/10.1007/BF01456804},
}

@article{anderson-1963-pca-asymptotics,
        ISSN = {00034851, 21688990},
        URL = {http://www.jstor.org/stable/2991288},
        author = {T. W. Anderson},
        journal = {The Annals of Mathematical Statistics},
        number = {1},
        pages = {122--148},
        publisher = {Institute of Mathematical Statistics},
        title = {Asymptotic Theory for Principal Component Analysis},
        urldate = {2025-09-14},
        volume = {34},
        year = {1963},
}

@book{jolliffe-2002-pca,
        author = {Ian T. Jolliffe},
        title = {Principal Component Analysis},
        edition = {2},
        year = {2002},
        publisher = {Springer},
        address = {New York / Heidelberg / Berlin},
        url = {
               http://cda.psych.uiuc.edu/statistical_learning_course/Jolliffe%20I.%20Principal%20Component%20Analysis%20(2ed.
               ,%20Springer,%202002)(518s)_MVsa_.pdf},
}

@article{waternaux-1976-nonnormal-eigenvalues,
        author = {Christine M. Waternaux},
        title = {Asymptotic distribution of the sample roots for a nonnormal
                 population},
        journal = {Biometrika},
        volume = {63},
        number = {3},
        pages = {639--645},
        year = {1976},
        doi = {10.1093/biomet/63.3.639},
        url = {https://academic.oup.com/biomet/article/63/3/639},
}

@article{eaton-1991-eigenvalues,
        author = {M. L. Eaton and D. E. Tyler},
        title = {On Wielandt’s Inequality and Its Application to the Asymptotic
                 Distribution of the Eigenvalues of a Random Symmetric Matrix},
        journal = {Annals of Statistics},
        volume = {19},
        number = {1},
        pages = {260--271},
        year = {1991},
        doi = {10.1214/aos/1176347980},
        url = {
               https://projecteuclid.org/journals/annals-of-statistics/volume-19/issue-1/On-Wielandts-Inequality-and-Its-Application-to-the-Asymptotic-Distribution/10.1214/aos/1176347980.full
               },
}

@book{zagidullina-2021-random-matrix-theory,
        author = {Aygul Zagidullina},
        title = {High-Dimensional Covariance Matrix Estimation: An Introduction
                 to Random Matrix Theory},
        series = {SpringerBriefs in Applied Statistics and Econometrics},
        year = {2021},
        publisher = {Springer Cham},
        doi = {10.1007/978-3-030-80065-9},
        url = {https://link.springer.com/book/10.1007/978-3-030-80065-9},
        pages = {115},
}

@book{anderson-2003-multivariate,
        author = {Anderson, T. W.},
        title = {An Introduction to Multivariate Statistical Analysis},
        edition = {3rd},
        publisher = {Wiley},
        address = {Hoboken, NJ},
        year = {2003},
        url = {
               https://ia903105.us.archive.org/28/items/zashu/AnIntroductionToMultivariateStatisticalAnalysis.pdf
               },
}

@book{kato-1980-perturbation,
        author = {Kato, Tosio},
        title = {Perturbation Theory for Linear Operators},
        edition = {Corrected printing of the second edition},
        series = {Grundlehren der Mathematischen Wissenschaften},
        volume = {132},
        publisher = {Springer-Verlag},
        address = {Berlin Heidelberg New York},
        year = {1980},
        url = {https://webhomes.maths.ed.ac.uk/~v1ranick/papers/kato1.pdf},
}

@article{abramowitz-saunders-2008-polarization-myth,
        author = {Abramowitz, Alan I. and Saunders, Kyle L.},
        title = {Is Polarization a Myth?},
        journal = {The Journal of Politics},
        year = {2008},
        volume = {70},
        number = {2},
        pages = {542--555},
}

@book{fiorina-abrams-2011-culture-war,
        author = {Fiorina, Morris P. and Abrams, Samuel J. and Pope, Jeremy C.},
        title = {Culture War? The Myth of a Polarized America},
        edition = {3},
        year = {2011},
        publisher = {Longman},
}

@article{iyengar-lelkes-2019-affective-polarization,
        author = {Iyengar, Shanto and Lelkes, Yphtach and Levendusky, Matthew
                  and Malhotra, Neil and Westwood, Sean J.},
        title = {The Origins and Consequences of Affective Polarization in the
                 United States},
        journal = {Annual Review of Political Science},
        year = {2019},
        volume = {22},
        pages = {129--146},
}

@article{lelkes-sood-iyengar-2017-hostile-audience,
        author = {Lelkes, Yphtach and Sood, Gaurav and Iyengar, Shanto},
        title = {The Hostile Audience: The Effect of Access to Broadband
                 Internet on Partisan Affect},
        journal = {American Journal of Political Science},
        year = {2017},
        volume = {61},
        number = {1},
        pages = {5--20},
        doi = {10.1111/ajps.12237},
        url = {https://doi.org/10.1111/ajps.12237},
}

@article{iyengar-sood-lelkes-2012-affect,
        author = {Iyengar, Shanto and Sood, Gaurav and Lelkes, Yphtach},
        title = {Affect, Not Ideology: A Social Identity Perspective on
                 Polarization},
        journal = {Public Opinion Quarterly},
        year = {2012},
        volume = {76},
        number = {3},
        pages = {405--431},
        month = {9},
        doi = {10.1093/poq/nfs038},
        url = {https://doi.org/10.1093/poq/nfs038},
}

@article{boxell-gentzkow-2017-internet-polarization,
        author = {Boxell, Levi and Gentzkow, Matthew and Shapiro, Jesse M.},
        title = {Greater Internet Use Is Not Associated with Faster Growth in
                 Political Polarization among U.S. Public},
        journal = {Proceedings of the National Academy of Sciences},
        year = {2017},
        volume = {114},
        number = {40},
        pages = {10612--10617},
}

@article{boxell-gentzkow-2023-cross-country-affective,
        author = {Boxell, Levi and Gentzkow, Matthew and Shapiro, Jesse M.},
        title = {Cross-Country Trends in Affective Polarization},
        journal = {The Review of Economics and Statistics},
        year = {2023},
}

@article{gennaioli-tabellini-2021-identity-beliefs,
        author = {Gennaioli, Nicola and Tabellini, Guido},
        title = {Identity, Beliefs, and Political Conflict},
        journal = {The Quarterly Journal of Economics},
        year = {2021},
        volume = {136},
        number = {4},
        pages = {2371--2411},
}

@book{drutman-2020-doom-loop,
        author = {Drutman, Lee},
        title = {Breaking the Two-Party Doom Loop: The Case for Multiparty
                 Democracy in America},
        year = {2020},
        publisher = {Oxford University Press},
}

@article{elsas-fiselier-2023-elite-polarization-dimensions,
        author = {van Elsas, Erika and Fiselier, Toine},
        title = {Conflict or Choice? The Differential Effects of Elite
                 Incivility and Ideological Polarization on Political Support},
        journal = {Acta Politica},
        year = {2023},
        volume = {59},
        pages = {589--618},
}

@article{knoll-2024-elite-polarization-boon-bane,
        author = {Knoll, Sabine and van den Bos, Karin},
        title = {Elite polarization — The boon and bane of democracy: Evidence
                 from Western Europe},
        journal = {Electoral Studies},
        year = {2024},
}

@book{mccarty-poole-rosenthal-2006-polarized-america,
        author = {McCarty, Nolan and Poole, Keith T. and Rosenthal, Howard},
        title = {Polarized America: The Dance of Ideology and Unequal Riches},
        publisher = {MIT Press},
        address = {Cambridge, MA},
        year = {2006},
}

@article{ojer-etal-2025-multidimensional-polarization,
        author = {Ojer, Jaume and C{\'a}rcamo, David and Pastor-Satorras,
                  Romualdo and Starnini, Michele},
        title = {Charting multidimensional ideological polarization across
                 demographic groups in the USA},
        journal = {Nature Human Behaviour},
        year = {2025},
}

@article{lelkes-2016-mass-elite-polarization,
        author = {Yphtach Lelkes},
        title = {Mass Polarization: Manifestations and Measurements},
        journal = {Public Opinion Quarterly},
        year = {2016},
        volume = {80},
        number = {S1},
        pages = {392--410},
        doi = {10.1093/poq/nfw005},
        url = {https://academic.oup.com/poq/article/80/S1/392/2223374},
}

@article{autor-dorn-hanson-majlesi-2020-importing-political-polarization,
        title = {Importing Political Polarization? The Electoral Consequences of
                 Rising Trade Exposure},
        author = {Autor, David H. and Dorn, David and Hanson, Gordon H. and
                  Majlesi, Kaveh},
        journal = {American Economic Review},
        volume = {110},
        number = {10},
        pages = {3139--3183},
        month = {10},
        year = {2020},
        doi = {10.1257/aer.20170011},
        url = {
               https://economics.mit.edu/sites/default/files/publications/Autor%20et%20al.%20-%202020%20-%20Importing%20Political%20Polariza.pdf
               },
}

@article{abramowitz-webster-2016-negative-partisanship,
        author = {Alan I. Abramowitz and Steven W. Webster},
        title = {The Rise of Negative Partisanship and the Nationalization of
                 U.S. Elections in the 21st Century},
        journal = {Electoral Studies},
        volume = {41},
        pages = {12--22},
        year = {2016},
        month = {3},
        doi = {10.1016/j.electstud.2015.11.001},
        url = {
               https://www.sciencedirect.com/science/article/pii/S0261379415001857
               },
        issn = {0261-3794},
}

@article{glaeser-ward-2006-american-political-geography,
        author = {Glaeser, Edward L. and Ward, Bryce A.},
        title = {Myths and Realities of American Political Geography},
        journal = {Journal of Economic Perspectives},
        volume = {20},
        number = {2},
        pages = {119--144},
        year = {2006},
        doi = {10.1257/jep.20.2.119},
        url = {https://www.aeaweb.org/articles?id=10.1257/jep.20.2.119},
}

\newpage

\appendix

\section{GSS Question Categorization}
\label{sec:gss-questions}

\subsection{Abortion}
    \textsc{abany} \hfill \textit{(Abortion if woman wants for any reason)}
    \begin{quote}
        \footnotesize
        Please tell me whether or not you think it should be possible for a pregnant woman to obtain a legal abortion if$\ldots$ READ EACH STATEMENT, AND CIRCLE ONE CODE FOR EACH. \\G. The woman wants it for any reason? 

        \textit{Possible responses}: ``yes'' (1), ``no'' (2)
    \end{quote}

    \textsc{abdefect} \hfill \textit{(Strong chance of serious defect)}
    \begin{quote}
        \footnotesize
        Please tell me whether or not you think it should be possible for a pregnant woman to obtain a legal abortion if$\ldots$ READ EACH STATEMENT, AND CIRCLE ONE CODE FOR EACH. \\A. If there is a strong chance of serious defect in the baby? 

        \textit{Possible responses}: ``yes'' (1), ``no'' (2)
    \end{quote}

    \textsc{abhlth} \hfill \textit{(Woman's health seriously endangered)}
    \begin{quote}
        \footnotesize
        Please tell me whether or not you think it should be possible for a pregnant woman to obtain a legal abortion if$\ldots$ READ EACH STATEMENT, AND CIRCLE ONE CODE FOR EACH. \\C. If the woman''s own health is seriously endangered by the pregnancy? 

        \textit{Possible responses}: ``yes'' (1), ``no'' (2)
    \end{quote}

    \textsc{abnomore} \hfill \textit{(Married--wants no more children)}
    \begin{quote}
        \footnotesize
        Please tell me whether or not you think it should be possible for a pregnant woman to obtain a legal abortion if$\ldots$ READ EACH STATEMENT, AND CIRCLE ONE CODE FOR EACH. \\B. If she is married and does not want any more children? 

        \textit{Possible responses}: ``yes'' (1), ``no'' (2)
    \end{quote}

    \textsc{abpoor} \hfill \textit{(Low income--cant afford more children)}
    \begin{quote}
        \footnotesize
        Please tell me whether or not you think it should be possible for a pregnant woman to obtain a legal abortion if$\ldots$ READ EACH STATEMENT, AND CIRCLE ONE CODE FOR EACH. \\D. If the family has a very low income and cannot afford any more children? 

        \textit{Possible responses}: ``yes'' (1), ``no'' (2)
    \end{quote}

    \textsc{abrape} \hfill \textit{(Pregnant as result of rape)}
    \begin{quote}
        \footnotesize
        Please tell me whether or not you think it should be possible for a pregnant woman to obtain a legal abortion if$\ldots$ READ EACH STATEMENT, AND CIRCLE ONE CODE FOR EACH. \\E. If she became pregnant as a result of rape? 

        \textit{Possible responses}: ``yes'' (1), ``no'' (2)
    \end{quote}

    \textsc{absingle} \hfill \textit{(Not married)}
    \begin{quote}
        \footnotesize
        Please tell me whether or not you think it should be possible for a pregnant woman to obtain a legal abortion if$\ldots$ READ EACH STATEMENT, AND CIRCLE ONE CODE FOR EACH. \\F. If she is not married and does not want to marry the man? 

        \textit{Possible responses}: ``yes'' (1), ``no'' (2)
    \end{quote}
    
\subsection{Affirmative Action}
    \textsc{affrmact} \hfill \textit{(Favor preference in hiring black people)}
    \begin{quote}
        \footnotesize
        A. Some people say that because of past discrimination, Black people should be given preference in hiring and promotion. Others say that such preference in hiring and promotion of Black people is wrong because it discriminates against whites. What about your opinion -- are you for or against preferential hiring and promotion of Black people? IF FAVORS: A. Do you favor preference in hiring and promotion strongly or not strongly? IF OPPOSES: B. Do you oppose preference in hiring and promotion strongly or not strongly? 

        \textit{Possible responses}: ``strongly favors'' (1), ``not strongly favors'' (2), ``not strongly opposes'' (3), ``strongly opposes'' (4)
    \end{quote}

    \textsc{fehire} \hfill \textit{(Should hire and promote women)}
    \begin{quote}
        \footnotesize
        Now I'm going to read several statements. As I read each one, please tell me whether you strongly agree, agree, neither agree nor disagree, disagree, or strongly disagree. For example, here is the statement: \\Because of past discrimination, employers should make special efforts to hire and promote qualified women. 

        \textit{Possible responses}: ``strongly agree'' (1), ``agree'' (2), ``neither agree nor disagree'' (3), ``disagree'' (4), ``strongly disagree'' (5)
    \end{quote}

    \textsc{fejobaff} \hfill \textit{(For or against preferential hiring of women)}
    \begin{quote}
        \footnotesize
        Some people say that because of past discrimination, women should be given preference in hiring and promotion. Others say that such preference in hiring and promotion of women is wrong because it discriminates against men. What about your opinion - are you for or against preferential hiring and promotion of women?\\IF FOR:Do you favor preference in hiring and promotion strongly or not strongly?\\IF AGAINST:Do you oppose preference in hiring and promotion strongly or not strongly? 

        \textit{Possible responses}: ``strongly favor'' (1), ``not strongly favor'' (2), ``not strongly oppose'' (3), ``strongly oppose'' (4)
    \end{quote}

    \textsc{helpblk} \hfill \textit{(Should govt aid black people?)}
    \begin{quote}
        \footnotesize
        Now look at CARD BF. Some people think that Black people have been discriminated against for so long that the government has a special obligation to help improve their living standards. Others believe that the government should not be giving special treatment to Black people.A. Where would you place yourself on this scale, or haven't you made up your mind on this? 

        \textit{Possible responses}: ``government should help'' (1), ``agree with both'' (3), ``no special treatment'' (5)
    \end{quote}
    
\subsection{Free Speech}
    \textsc{colath} \hfill \textit{(Allow anti-religionist to teach)}
    \begin{quote}
        \footnotesize
        There are always some people whose ideas are considered bad or dangerous by other people. For instance, somebody who is against all churches and religion $\ldots$ \\B. Should such a person be allowed to teach in a college or university, or not? 

        \textit{Possible responses}: ``yes, allowed to teach'' (4), ``not allowed'' (5)
    \end{quote}

    \textsc{colhomo} \hfill \textit{(Allow homosexual to teach)}
    \begin{quote}
        \footnotesize
        And what about a man who admits that he is a homosexual? \\B. Should such a person be allowed to teach in a college or university, or not? 

        \textit{Possible responses}: ``yes, allowed to teach'' (4), ``not allowed'' (5)
    \end{quote}

    \textsc{colmil} \hfill \textit{(Allow militarist to teach)}
    \begin{quote}
        \footnotesize
        Consider a person who advocates doing away with elections and letting the military run the country.\\B. Should such a person be allowed to teach in a college or university, or not? 

        \textit{Possible responses}: ``yes, allowed to teach'' (4), ``not allowed'' (5)
    \end{quote}

    \textsc{colmslm} \hfill \textit{(Allow anti-american muslim clergymen teaching in college)}
    \begin{quote}
        \footnotesize
        Now consider a Muslim clergyman who preaches hatred of the United States. \\B. Should such a person be allowed to teach in a college or university, or not? 

        \textit{Possible responses}: ``yes, allowed to teach'' (4), ``not allowed'' (5)
    \end{quote}

    \textsc{colrac} \hfill \textit{(Allow racist to teach)}
    \begin{quote}
        \footnotesize
        Or consider a person who believes that Black people are genetically inferior....B. Should such a person be allowed to teach in a college or university, or not? 

        \textit{Possible responses}: ``yes, allowed to teach'' (4), ``not allowed'' (5)
    \end{quote}

    \textsc{libath} \hfill \textit{(Allow anti-religious book in library)}
    \begin{quote}
        \footnotesize
        There are always some people whose ideas are considered bad or dangerous by other people. For instance, somebody who is against all churches and religion $\ldots$ \\C. If some people in your community suggested that a book he wrote against churches and religion should be taken out of your public library, would you favor removing this book, or not? 

        \textit{Possible responses}: ``remove'' (1), ``not remove'' (2)
    \end{quote}

    \textsc{libhomo} \hfill \textit{(Allow homosexuals book in library)}
    \begin{quote}
        \footnotesize
        And what about a man who admits that he is a homosexual? \\C. If some people in your community suggested that a book he wrote in favor of homosexuality should be taken out of your public library, would you favor removing this book, or not? 

        \textit{Possible responses}: ``remove'' (1), ``not remove'' (2)
    \end{quote}

    \textsc{libmil} \hfill \textit{(Allow militarists book in library)}
    \begin{quote}
        \footnotesize
        Consider a person who advocates doing away with elections and letting the military run the country.\\C. Suppose he wrote a book advocating doing away with elections and letting the military run the country. somebody in your community suggests that the book be removed from the public library. Would you favor removing it, or not? 

        \textit{Possible responses}: ``remove'' (1), ``not remove'' (2)
    \end{quote}

    \textsc{libmslm} \hfill \textit{(Allow anti-american muslim clergymen's books in library)}
    \begin{quote}
        \footnotesize
        Now consider a Muslim clergyman who preaches hatred of the United States.\\C. If some people in your community suggested that a book he wrote which preaches hatred of the United States should be taken out of your public library, would you favor removing this book, or not? 

        \textit{Possible responses}: ``remove'' (1), ``not remove'' (2)
    \end{quote}

    \textsc{librac} \hfill \textit{(Allow racists book in library)}
    \begin{quote}
        \footnotesize
        Or consider a person who believes that Black people are genetically inferior.C. If some people in your community suggested that a book he wrote which said Black people are inferior should be taken out of your public library, would you favor removing this book, or not? 

        \textit{Possible responses}: ``remove'' (1), ``not remove'' (2)
    \end{quote}

    \textsc{spkath} \hfill \textit{(Allow anti-religionist to speak)}
    \begin{quote}
        \footnotesize
        There are always some people whose ideas are considered bad or dangerous by other people. For instance, somebody who is against all churches and religion $\ldots$ \\A. If such a person wanted to make a speech in your (city/town/community) against churches and religion, should he be allowed to speak, or not? 

        \textit{Possible responses}: ``yes, allowed to speak'' (1), ``not allowed'' (2)
    \end{quote}

    \textsc{spkhomo} \hfill \textit{(Allow homosexual to speak)}
    \begin{quote}
        \footnotesize
        And what about a man who admits that he is a homosexual? \\A. Suppose this admitted homosexual wanted to make a speech in your community. Should he be allowed to speak, or not? 

        \textit{Possible responses}: ``yes, allowed to speak'' (1), ``not allowed'' (2)
    \end{quote}

    \textsc{spkmil} \hfill \textit{(Allow militarist to speak)}
    \begin{quote}
        \footnotesize
        Consider a person who advocates doing away with elections and letting the military run the country.\\A. If such a person wanted to make a speech in your community, should he be allowed to speak, or not? 

        \textit{Possible responses}: ``yes, allowed to speak'' (1), ``not allowed'' (2)
    \end{quote}

    \textsc{spkmslm} \hfill \textit{(Allow muslim clergymen preaching hatred of the us)}
    \begin{quote}
        \footnotesize
        Now consider a Muslim clergyman who preaches hatred of the United States. \\A. If such a person wanted to make a speech in your community preaching hatred of the United States, should he be allowed  to speak, or not? 

        \textit{Possible responses}: ``yes, allowed to speak'' (1), ``not allowed'' (2)
    \end{quote}

    \textsc{spkrac} \hfill \textit{(Allow racist to speak)}
    \begin{quote}
        \footnotesize
        Or consider a person who believes that Black people are genetically inferior... A. If such a person wanted to make a speech in your community claiming that Black people are inferior, should he be allowed to speak, or not? 

        \textit{Possible responses}: ``yes, allowed to speak'' (1), ``not allowed'' (2)
    \end{quote}
    
\subsection{Government Spending}
    \textsc{advfront} \hfill \textit{(Sci rsch is necessary and should be supported by federal govt)}
    \begin{quote}
        \footnotesize
        I’m going to read to you some statements like those you might find in a newspaper or magazine article. For each statement, please tell me if you strongly agree, agree, disagree, or strongly disagree.\\D. Even if it brings no immediate benefits, scientific research that advances the frontiers of knowledge is necessary and should be supported by the federal government. 

        \textit{Possible responses}: ``strongly agree'' (1), ``agree'' (2), ``disagree'' (3), ``strongly disagree'' (4)
    \end{quote}

    \textsc{nataid} \hfill \textit{(Foreign aid)}
    \begin{quote}
        \footnotesize
        We are faced with many problems in this country, none of which can be solved easily or inexpensively. I'm going to name some of these problems, and for each one I'd like you to name some of these problems, and for each one I'd like you to tell me whether you think we're spending too much money on it, too little money, or about the right amount. First (READ ITEM A) $\ldots$  are we spending too much, too little, or about the right amount on (ITEM)?\\J. Foreign aid 

        \textit{Possible responses}: ``too little'' (1), ``about right'' (2), ``too much'' (3)
    \end{quote}

    \textsc{natarms} \hfill \textit{(Military, armaments, and defense)}
    \begin{quote}
        \footnotesize
        We are faced with many problems in this country, none of which can be solved easily or inexpensively. I'm going to name some of these problems, and for each one I'd like you to name some of these problems, and for each one I'd like you to tell me whether you think we're spending too much money on it, too little money, or about the right amount. First (READ ITEM A) $\ldots$  are we spending too much, too little, or about the right amount on (ITEM)?\\I. The military, armaments and defense 

        \textit{Possible responses}: ``too little'' (1), ``about right'' (2), ``too much'' (3)
    \end{quote}

    \textsc{natcity} \hfill \textit{(Solving problems of big cities)}
    \begin{quote}
        \footnotesize
        We are faced with many problems in this country, none of which can be solved easily or inexpensively. I'm going to name some of these problems, and for each one I'd like you to name some of these problems, and for each one I'd like you to tell me whether you think we're spending too much money on it, too little money, or about the right amount. First (READ ITEM A) $\ldots$  are we spending too much, too little, or about the right amount on (ITEM)?\\D. Solving the problems of the big cities 

        \textit{Possible responses}: ``too little'' (1), ``about right'' (2), ``too much'' (3)
    \end{quote}

    \textsc{natcrime} \hfill \textit{(Halting rising crime rate)}
    \begin{quote}
        \footnotesize
        We are faced with many problems in this country, none of which can be solved easily or inexpensively. I'm going to name some of these problems, and for each one I'd like you to name some of these problems, and for each one I'd like you to tell me whether you think we're spending too much money on it, too little money, or about the right amount. First (READ ITEM A) $\ldots$  are we spending too much, too little, or about the right amount on (ITEM)?\\E. Halting the rising crime rate 

        \textit{Possible responses}: ``too little'' (1), ``about right'' (2), ``too much'' (3)
    \end{quote}

    \textsc{natdrug} \hfill \textit{(Dealing with drug addiction)}
    \begin{quote}
        \footnotesize
        We are faced with many problems in this country, none of which can be solved easily or inexpensively. I'm going to name some of these problems, and for each one I'd like you to name some of these problems, and for each one I'd like you to tell me whether you think we're spending too much money on it, too little money, or about the right amount. First (READ ITEM A) $\ldots$  are we spending too much, too little, or about the right amount on (ITEM)?\\F. Dealing with drug addiction 

        \textit{Possible responses}: ``too little'' (1), ``about right'' (2), ``too much'' (3)
    \end{quote}

    \textsc{nateduc} \hfill \textit{(Improving nations education system)}
    \begin{quote}
        \footnotesize
        We are faced with many problems in this country, none of which can be solved easily or inexpensively. I'm going to name some of these problems, and for each one I'd like you to name some of these problems, and for each one I'd like you to tell me whether you think we're spending too much money on it, too little money, or about the right amount. First (READ ITEM A) $\ldots$  are we spending too much, too little, or about the right amount on (ITEM)?\\G. Improving the nation's education system 

        \textit{Possible responses}: ``too little'' (1), ``about right'' (2), ``too much'' (3)
    \end{quote}

    \textsc{natenvir} \hfill \textit{(Improving \& protecting environment)}
    \begin{quote}
        \footnotesize
        We are faced with many problems in this country, none of which can be solved easily or inexpensively. I'm going to name some of these problems, and for each one I'd like you to name some of these problems, and for each one I'd like you to tell me whether you think we're spending too much money on it, too little money, or about the right amount. First (READ ITEM A) $\ldots$  are we spending too much, too little, or about the right amount on (ITEM)?\\B. Improving and protecting the environment 

        \textit{Possible responses}: ``too little'' (1), ``about right'' (2), ``too much'' (3)
    \end{quote}

    \textsc{natfare} \hfill \textit{(Welfare)}
    \begin{quote}
        \footnotesize
        We are faced with many problems in this country, none of which can be solved easily or inexpensively. I'm going to name some of these problems, and for each one I'd like you to name some of these problems, and for each one I'd like you to tell me whether you think we're spending too much money on it, too little money, or about the right amount. First (READ ITEM A) $\ldots$  are we spending too much, too little, or about the right amount on (ITEM)?\\K. Welfare 

        \textit{Possible responses}: ``too little'' (1), ``about right'' (2), ``too much'' (3)
    \end{quote}

    \textsc{natheal} \hfill \textit{(Improving \& protecting nations health)}
    \begin{quote}
        \footnotesize
        We are faced with many problems in this country, none of which can be solved easily or inexpensively. I'm going to name some of these problems, and for each one I'd like you to name some of these problems, and for each one I'd like you to tell me whether you think we're spending too much money on it, too little money, or about the right amount. First (READ ITEM A) $\ldots$  are we spending too much, too little, or about the right amount on (ITEM)?\\C. Improving and protecting the nation's health 

        \textit{Possible responses}: ``too little'' (1), ``about right'' (2), ``too much'' (3)
    \end{quote}

    \textsc{natmass} \hfill \textit{(Mass transportation)}
    \begin{quote}
        \footnotesize
        We are faced with many problems in this country, none of which can be solved easily or inexpensively. I'm going to name some of these problems, and for each one I'd like you to name some of these problems, and for each one I'd like you to tell me whether you think we're spending too much money on it, too little money, or about the right amount. First (READ ITEM A) $\ldots$  are we spending too much, too little, or about the right amount on (ITEM)?\\N. Mass Transportation 

        \textit{Possible responses}: ``too little'' (1), ``about right'' (2), ``too much'' (3)
    \end{quote}

    \textsc{natpark} \hfill \textit{(Parks and recreation)}
    \begin{quote}
        \footnotesize
        We are faced with many problems in this country, none of which can be solved easily or inexpensively. I'm going to name some of these problems, and for each one I'd like you to name some of these problems, and for each one I'd like you to tell me whether you think we're spending too much money on it, too little money, or about the right amount. First (READ ITEM A) $\ldots$  are we spending too much, too little, or about the right amount on (ITEM)?\\O. Parks and recreation 

        \textit{Possible responses}: ``too little'' (1), ``about right'' (2), ``too much'' (3)
    \end{quote}

    \textsc{natrace} \hfill \textit{(Improving the conditions of black people)}
    \begin{quote}
        \footnotesize
        We are faced with many problems in this country, none of which can be solved easily or inexpensively. I'm going to name some of these problems, and for each one I'd like you to name some of these problems, and for each one I'd like you to tell me whether you think we're spending too much money on it, too little money, or about the right amount. First (READ ITEM A) $\ldots$  are we spending too much, too little, or about the right amount on (ITEM)?H. Improving the conditions of Black people. 

        \textit{Possible responses}: ``too little'' (1), ``about right'' (2), ``too much'' (3)
    \end{quote}

    \textsc{natroad} \hfill \textit{(Highways and bridges)}
    \begin{quote}
        \footnotesize
        We are faced with many problems in this country, none of which can be solved easily or inexpensively. I'm going to name some of these problems, and for each one I'd like you to name some of these problems, and for each one I'd like you to tell me whether you think we're spending too much money on it, too little money, or about the right amount. First (READ ITEM A) $\ldots$  are we spending too much, too little, or about the right amount on (ITEM)?\\L. Highways and bridges 

        \textit{Possible responses}: ``too little'' (1), ``about right'' (2), ``too much'' (3)
    \end{quote}

    \textsc{natsoc} \hfill \textit{(Social security)}
    \begin{quote}
        \footnotesize
        We are faced with many problems in this country, none of which can be solved easily or inexpensively. I'm going to name some of these problems, and for each one I'd like you to name some of these problems, and for each one I'd like you to tell me whether you think we're spending too much money on it, too little money, or about the right amount. First (READ ITEM A) $\ldots$  are we spending too much, too little, or about the right amount on (ITEM)?\\M. Social Security 

        \textit{Possible responses}: ``too little'' (1), ``about right'' (2), ``too much'' (3)
    \end{quote}

    \textsc{natspac} \hfill \textit{(Space exploration program)}
    \begin{quote}
        \footnotesize
        We are faced with many problems in this country, none of which can be solved easily or inexpensively. I'm going to name some of these problems, and for each one I'd like you to name some of these problems, and for each one I'd like you to tell me whether you think we're spending too much money on it, too little money, or about the right amount. First (READ ITEM A) $\ldots$  are we spending too much, too little, or about the right amount on (ITEM)?\\A. Space exploration program 

        \textit{Possible responses}: ``too little'' (1), ``about right'' (2), ``too much'' (3)
    \end{quote}
    
\subsection{Police and the Judicial System}
    \textsc{cappun} \hfill \textit{(Favor or oppose death penalty for murder)}
    \begin{quote}
        \footnotesize
        Do you favor or oppose the death penalty for persons convicted of murder? 

        \textit{Possible responses}: ``favor'' (1), ``oppose'' (2)
    \end{quote}

    \textsc{conjudge} \hfill \textit{(Confid. in united states supreme court)}
    \begin{quote}
        \footnotesize
        I am going to name some institutions in this country. As far as the people running these institutions are concerned, would you say you have a great deal of confidence, only some confidence, or hardly any confidence at all in them?  \\READ EACH ITEM; CODE ONE FOR EACH.\\J. U.S. Supreme Court 

        \textit{Possible responses}: ``a great deal'' (1), ``only some'' (2), ``hardly any'' (3)
    \end{quote}

    \textsc{courts} \hfill \textit{(Courts dealing with criminals)}
    \begin{quote}
        \footnotesize
        In general, do you think the courts in this area deal too harshly or not harshly enough with criminals? 

        \textit{Possible responses}: ``too harshly'' (1), ``not harshly enough'' (2), ``about right'' (3)
    \end{quote}

    \textsc{natcrime} \hfill \textit{(Halting rising crime rate)}
    \begin{quote}
        \footnotesize
        We are faced with many problems in this country, none of which can be solved easily or inexpensively. I'm going to name some of these problems, and for each one I'd like you to name some of these problems, and for each one I'd like you to tell me whether you think we're spending too much money on it, too little money, or about the right amount. First (READ ITEM A) $\ldots$  are we spending too much, too little, or about the right amount on (ITEM)?\\E. Halting the rising crime rate 

        \textit{Possible responses}: ``too little'' (1), ``about right'' (2), ``too much'' (3)
    \end{quote}

    \textsc{polabuse} \hfill \textit{(Citizen said vulgar or obscene things)}
    \begin{quote}
        \footnotesize
        Are there any situations you can imagine in which you would approve of a policeman striking an adult male citizen? \\A. Had said vulgar and obscene things to the policeman? 

        \textit{Possible responses}: ``yes'' (1), ``no'' (2)
    \end{quote}

    \textsc{polattak} \hfill \textit{(Citizen attacking policeman with fists)}
    \begin{quote}
        \footnotesize
        Are there any situations you can imagine in which you would approve of a policeman striking an adult male citizen? \\D. Was attacking the policeman with his fists? 

        \textit{Possible responses}: ``yes'' (1), ``no'' (2)
    \end{quote}

    \textsc{polescap} \hfill \textit{(Citizen attempting to escape custody)}
    \begin{quote}
        \footnotesize
        Are there any situations you can imagine in which you would approve of a policeman striking an adult male citizen? \\C. Was attempting to escape from custody? 

        \textit{Possible responses}: ``yes'' (1), ``no'' (2)
    \end{quote}

    \textsc{polhitok} \hfill \textit{(Ever approve of police striking citizen)}
    \begin{quote}
        \footnotesize
        Are there any situations you can imagine in which you would approve of a policeman striking an adult male citizen? 

        \textit{Possible responses}: ``yes'' (1), ``no'' (2)
    \end{quote}

    \textsc{polmurdr} \hfill \textit{(Citizen questioned as murder suspect)}
    \begin{quote}
        \footnotesize
        Are there any situations you can imagine in which you would approve of a policeman striking an adult male citizen? \\B. Was being questioned as a suspect in a murder case? 

        \textit{Possible responses}: ``yes'' (1), ``no'' (2)
    \end{quote}
    
\subsection{Race}
    \textsc{colrac} \hfill \textit{(Allow racist to teach)}
    \begin{quote}
        \footnotesize
        Or consider a person who believes that Black people are genetically inferior....B. Should such a person be allowed to teach in a college or university, or not? 

        \textit{Possible responses}: ``yes, allowed to teach'' (4), ``not allowed'' (5)
    \end{quote}

    \textsc{librac} \hfill \textit{(Allow racists book in library)}
    \begin{quote}
        \footnotesize
        Or consider a person who believes that Black people are genetically inferior.C. If some people in your community suggested that a book he wrote which said Black people are inferior should be taken out of your public library, would you favor removing this book, or not? 

        \textit{Possible responses}: ``remove'' (1), ``not remove'' (2)
    \end{quote}

    \textsc{natrace} \hfill \textit{(Improving the conditions of black people)}
    \begin{quote}
        \footnotesize
        We are faced with many problems in this country, none of which can be solved easily or inexpensively. I'm going to name some of these problems, and for each one I'd like you to name some of these problems, and for each one I'd like you to tell me whether you think we're spending too much money on it, too little money, or about the right amount. First (READ ITEM A) $\ldots$  are we spending too much, too little, or about the right amount on (ITEM)?H. Improving the conditions of Black people. 

        \textit{Possible responses}: ``too little'' (1), ``about right'' (2), ``too much'' (3)
    \end{quote}

    \textsc{racdif1} \hfill \textit{(Differences due to discrimination)}
    \begin{quote}
        \footnotesize
        On the average Black people have worse jobs, income, and housing than white people. Do you think these differences are $\ldots$  A. Mainly due to discrimination? 

        \textit{Possible responses}: ``yes'' (1), ``no'' (2)
    \end{quote}

    \textsc{racdif2} \hfill \textit{(Differences due to in-born learning ability)}
    \begin{quote}
        \footnotesize
        On the average Black people have worse jobs, income, and housing than white people. Do you think these differences are $\ldots$  B. Because most Black people have less in-born ability to learn? 

        \textit{Possible responses}: ``yes'' (1), ``no'' (2)
    \end{quote}

    \textsc{racdif3} \hfill \textit{(Differences due to lack of education)}
    \begin{quote}
        \footnotesize
        On the average Black people have worse jobs, income, and housing than white people.Do you think these differences are $\ldots$ C. Because most Black people don't have the chance for education that it takes to rise out of poverty? 

        \textit{Possible responses}: ``yes'' (1), ``no'' (2)
    \end{quote}

    \textsc{racdif4} \hfill \textit{(Differences due to lack of will)}
    \begin{quote}
        \footnotesize
        On the average Black people have worse jobs, income, and housing than white people.Do you think these differences are $\ldots$ D. Because most Black people just don't have the motivation or will power to pull themselves up out of poverty? 

        \textit{Possible responses}: ``yes'' (1), ``no'' (2)
    \end{quote}

    \textsc{spkrac} \hfill \textit{(Allow racist to speak)}
    \begin{quote}
        \footnotesize
        Or consider a person who believes that Black people are genetically inferior... A. If such a person wanted to make a speech in your community claiming that Black people are inferior, should he be allowed to speak, or not? 

        \textit{Possible responses}: ``yes, allowed to speak'' (1), ``not allowed'' (2)
    \end{quote}

    \textsc{wrkwayup} \hfill \textit{(Black people overcome prejudice without favors)}
    \begin{quote}
        \footnotesize
        B. Do you agree strongly, agree somewhat, neither agree nor disagree, disagree somewhat, or disagree strongly with the following statement (HAND CARD TO RESPONDENT): Irish, Italians, Jewish and many other minorities overcame prejudice and worked their way up. Black people should do the same without special favors. 

        \textit{Possible responses}: ``agree strongly'' (1), ``agree somewhat'' (2), ``neither agree nor disagree'' (3), ``disagree somewhat'' (4), ``disagree strongly'' (5)
    \end{quote}
    
\subsection{Sex and Birth Control}
    \textsc{homosex} \hfill \textit{(Homosexual sex relations)}
    \begin{quote}
        \footnotesize
        What about sexual relations between two adults of the same sex--do you think it is always wrong, almost always wrong, wrong only sometimes, or not wrong at all? 

        \textit{Possible responses}: ``always wrong'' (1), ``almost always wrong'' (2), ``wrong only sometimes'' (3), ``not wrong at all'' (4), ``other'' (5)
    \end{quote}

    \textsc{pillok} \hfill \textit{(Birth control to teenagers 14-16)}
    \begin{quote}
        \footnotesize
        C. Do you strongly agree, agree, disagree, or strongly disagree that methods of birth control should be available to teenagers between the ages of 14 and 16 if their parents do not approve? 

        \textit{Possible responses}: ``strongly agree'' (1), ``agree'' (2), ``disagree'' (3), ``strongly disagree'' (4)
    \end{quote}

    \textsc{premarsx} \hfill \textit{(Sex before marriage)}
    \begin{quote}
        \footnotesize
        There's been a lot of discussion about the way morals and attitudes about sex are changing in this country. If a man and woman have sex relations before marriage, do you think it is always wrong, almost always wrong, wrong only \\sometimes, or not wrong at all? 

        \textit{Possible responses}: ``always wrong'' (1), ``almost always wrong'' (2), ``wrong only sometimes'' (3), ``not wrong at all'' (4), ``other'' (5)
    \end{quote}

    \textsc{sexeduc} \hfill \textit{(Sex education in public schools)}
    \begin{quote}
        \footnotesize
        Would you be for or against sex education in the public schools? 

        \textit{Possible responses}: ``favor'' (1), ``oppose'' (2), ``depends on age/grade (vol.)'' (3)
    \end{quote}

    \textsc{teensex} \hfill \textit{(Sex before marriage:teens 14-16)}
    \begin{quote}
        \footnotesize
        There's been a lot of discussion about the way morals and attitudes about sex are changing in this country. If a man and woman have sex relations before marriage, do you think it is always wrong, almost always wrong, wrong only sometimes, or not wrong at all?\\A. What if they are in their early teens, say 14 to 16 years old?  In that case, do you think sex relations before marriage are always wrong, almost always wrong, wrong only sometimes, or not wrong at all? 

        \textit{Possible responses}: ``always wrong'' (1), ``almost always wrong'' (2), ``wrong only sometimes'' (3), ``not wrong at all'' (4), ``other'' (5)
    \end{quote}

    \textsc{xmarsex} \hfill \textit{(Sex with person other than spouse)}
    \begin{quote}
        \footnotesize
        What is your opinion about a married person having sexual relations with someone other than the marriage partner--is it always wrong, almost always wrong, wrong only sometimes, or not wrong at all? 

        \textit{Possible responses}: ``always wrong'' (1), ``almost always wrong'' (2), ``wrong only sometimes'' (3), ``not wrong at all'' (4), ``other'' (5)
    \end{quote}
    
\subsection{Welfare}
    \textsc{eqwlth} \hfill \textit{(Should govt reduce income differences)}
    \begin{quote}
        \footnotesize
        A. Some people think that the government in Washington ought to reduce the income differences between the rich and the poor, perhaps by raising the taxes of wealthy families or by giving income assistance to the poor. Others think that the government should not concern itself with reducing this income difference between the rich and the poor. Here is a card with a scale from 1 to 7. Think of a score of 1 as meaning that the government ought to reduce the income differences between rich and poor, and a score of 7 meaning that the government should not concern itself with reducing income differences. What score between 1 and 7 comes closest to the way you feel? (CIRCLE ONE): 

        \textit{Possible responses}: ``the government should reduce income differences'' (1), ``the government should not concern itself with reducing income differences'' (7)
    \end{quote}

    \textsc{helppoor} \hfill \textit{(Should govt improve standard of living?)}
    \begin{quote}
        \footnotesize
        I'd like to talk with you about issues some people tell us are important. Please look at CARD BC. Some people think that the government in Washington should do everything possible to improve the standard of living of all poor Americans; they are at Point 1 on this card. Other people think it is not the government's responsibility, and that each person should take care of himself; they are at Point 5.\\A. Where would you place yourself on this scale, or haven't you have up your mind on this? 

        \textit{Possible responses}: ``government should help'' (1), ``agree with both'' (3), ``people should care for themselves'' (5)
    \end{quote}

    \textsc{helpsick} \hfill \textit{(Should govt help pay for medical care?)}
    \begin{quote}
        \footnotesize
        Look at CARD BE. In general, some people think that it is the responsibility of the government in Washington to see to it that people have help in paying for doctors and hospital bills. Others think that these matters are not the responsibility of the federal government and that people should take care of these things themselves.\\A. Where would you place yourself on this scale, or haven't you made up your mind on this? 

        \textit{Possible responses}: ``government should help'' (1), ``agree with both'' (3), ``people should care for themselves'' (5)
    \end{quote}

    \textsc{natfare} \hfill \textit{(Welfare)}
    \begin{quote}
        \footnotesize
        We are faced with many problems in this country, none of which can be solved easily or inexpensively. I'm going to name some of these problems, and for each one I'd like you to name some of these problems, and for each one I'd like you to tell me whether you think we're spending too much money on it, too little money, or about the right amount. First (READ ITEM A) $\ldots$  are we spending too much, too little, or about the right amount on (ITEM)?\\K. Welfare 

        \textit{Possible responses}: ``too little'' (1), ``about right'' (2), ``too much'' (3)
    \end{quote}

\section{Deferred Proofs}
\label{sec:deferred-proofs}

{\color{red}\textbf{Note:} this section is under construction --- some proofs lack detail.}

\subsection{Spectral Radius Estimator}

Let $\lambda_1 \geq \lambda_2 \geq \cdots \geq \lambda_p \geq 0$ be the eigenvalues of $\Sigma$ and $\hat\lambda_1 \ge \hat\lambda_2 \ge \cdots \ge \hat\lambda_p$ the eigenvalues of $\widehat S_n$.

To do so, I leverage the fact that our matrix norm can be expressed in terms of the covariance eigenvalues and lean heavily on PCA theory going back to \textcite{anderson-1963-pca-asymptotics}.
For an overview, see \textcite{jolliffe-2002-pca} or \textcite{zagidullina-2021-random-matrix-theory}.
See \textcite{anderson-2003-multivariate} for more.

\begin{proposition}[Consistency of sample eigenvalues]
	\label{prop:sample-eigenvalue-consistency}
	The sample eigenvalues $\hat \lambda_i$ are consistent estimates of the population eigenvalues $\lambda_i$.
\end{proposition}

\begin{proof}
	I prove the stronger notion of almost-sure convergence. The (strong) law of large numbers gives almost-sure convergence of the covariance matrix $\widehat S_n \overset{a.s.}{\to} \Sigma$. The set of eigenvalues is a continuous function of a matrices entries \cite[Ch. 2, Th. 5.14, p. 118]{kato-1980-perturbation}, so the continuous mapping theorem yields that $\hat \lambda_i \overset{a.s.}{\to} \lambda_i$.
\end{proof}

\begin{proposition}[Asymptotic normality of sample eigenvalues]
	\label{prop:sample-eigenvalue-normality}
	Let $X_1,\dots,X_n \in \mathbb{R}^p$ be i.i.d.\ with mean $0$ and covariance $\Sigma$.
	If $\Sigma$ has eigenvalues $\lambda_1 \geq \lambda_2 \geq \cdots \geq \lambda_p \geq 0$ with corresponding sample eigenvalues $\hat\lambda_1 \ge \hat\lambda_2 \ge \cdots \ge \hat\lambda_p$, then
	\begin{equation}
		\sqrt{n} \paren{\hat\lambda_i - \lambda_i}
		\overset{d}{\longrightarrow}
		N\!\paren{0,\;\sigma_i^2},
		\qquad i=1,\dots,p.
	\end{equation}
	If $\set{X_i}$ are normally distributed, then $\sigma_i^2 = 2\lambda_i^2$. If not,
	$\sigma_i^2$ depends (somewhat complexly) on the fourth cumulants of $X_i$.
\end{proposition}

\begin{proof}
	The normal result is due to \textcite{anderson-1963-pca-asymptotics}.
	The non-normal case is due to \textcite{waternaux-1976-nonnormal-eigenvalues} and expanded upon by \textcite{eaton-1991-eigenvalues}.
	The non-normal case will be of most use to us because most survey data is decidedly not normal (e.g. binary response or multiple choice).
\end{proof}

Asymptotic normality and consistency of the spectral radius estimator follow trivially from \cref{prop:sample-eigenvalue-consistency} and \cref{prop:sample-eigenvalue-normality}.

\subsection{Latent Model}

I first establish a few lemmas that greatly simplify the proof.
The first is regarding the spectral radius of a positive rank-one update to a positive semidefinite matrix, which I prove as consequence of Weyl's inequalities \parencite{weyl-1912-inequalities,tao-2010-eigenvalue-blog}.
This result is very similar in flavor those of \textcite{golub-1973-eigenvalues}.
I then leverage this to prove our proposition and close with a discussion of two sufficient conditions under which the phrase "non-decreasing" on the last line of \cref{thm:one-dim-implies-multiple-dim} becomes "strictly increasing".

\begin{lemma}
	\label{lem:rank-one-update}
	If $D \in \R^{n \times n}$ is symmetric and positive semidefinite, $v \in \R^n$, and $c \geq 0$, then
	\begin{equation}
		\rho(c v v \trans + D) \geq \rho(D).
	\end{equation}
\end{lemma}

\begin{proof}
	See \textcite[\S 5]{golub-1973-eigenvalues}.
\end{proof}

\begin{lemma}
	\label{lem:rank-one-update-increasing}
	Let the setup be the same as in \cref{lem:rank-one-update}. Then $\rho(c v v \trans + D)$ is non-decreasing with respect to $c$.
\end{lemma}

\begin{proof}
	Let $\epsilon > 0$. Then,
	\begin{equation}
		\begin{aligned}
			\rho((c + \epsilon) v v \trans + D)
			\ = \ \rho(\epsilon v v \trans + (c v v \trans + D))
			\ \geq \ \rho(c v v \trans + D)
		\end{aligned}
	\end{equation}
	by \cref{lem:rank-one-update} because $c v v \trans + D \succeq 0$ by the closure of positive semidefinite matrices under addition.
\end{proof}

Proof of the original result is now quite simple:

\begin{proof}[Proof of \cref{thm:one-dim-implies-multiple-dim}]
	The spectral radius of $\Sigma$ is
	\begin{equation}
		\rho(\Sigma)
		\ = \ \rho(\Var(\beta x + \epsilon))
		\ = \ \rho(a \beta \beta \trans + \Gamma)
	\end{equation}
	Because $\Gamma$ is a covariance matrix, it's symmetric and positive semidefinite and \cref{lem:rank-one-update-increasing} gives us that $\rho(\sigma)$ is non-decreasing in $a$ as desired.
\end{proof}

Strict increase if $\beta$ nontrivially projects onto the first eigenspace of $\Gamma$. This happens on all but a measure zero subset of possible $(\beta, \Gamma)$ pairs.

\begin{proposition}
	If $a = \Var(x) > \rho(\Gamma) = r$, then $r$ is \emph{strictly increasing} with respect to $a$.
\end{proposition}

\begin{corollary}
	There exists some $a > 0$ after which $r$ is strictly increasing.
\end{corollary}

\section{Decompositions by Other Characteristics}
\label{sec:other-decompositions}

\foreach \name/\displayName in {
		sex/Sex,
		race/Race,
		religion/Religion,
		geographic_region/Geographic Region,
		type_of_neighborhood/Type of Neighborhood,
		political_ideology/Political Ideology,
		workforce_status/Workforce Status,
		self_or_parent_immigrant/Immigrant Status,
		education/Education,
		age_bucket/Age
	}{
		\begin{figure}
			\centering
			\caption{Polarization Within \displayName}
			\input{./figures/gss/by_group/\name.pgf}
		\end{figure}

		\begin{figure}
			\centering
			\caption{Polarization Trends Within vs. Between \displayName}
			\input{./figures/gss/decompositions/\name.pgf}
		\end{figure}
	}

\end{document}